\documentclass[epj,nopacs]{svjour}
\usepackage{graphicx}
\usepackage{color}
\usepackage{latexsym}
\usepackage[colorlinks,linkcolor=blue,citecolor=blue,urlcolor=black]{hyperref}
\usepackage{ulem}

\newcommand{\be}{\begin{eqnarray}}
\newcommand{\ee}{\end{eqnarray}}

\begin{document}
\title{Testing Ghasemi-Nodehi-Bambi metric parameters with quasi-periodic oscillations}
\author{M. Ghasemi-Nodehi\inst{1}\inst{2}\thanks{\emph{email:} mghasemin@ipm.ir}, Youjun Lu\inst{1}\inst{3}, Ju Chen\inst{1}\inst{3}, \& Chao Yang\inst{1}\inst{3}
%
}                     
%
%
\institute{ CAS Key laboratory for computational Astrophysics, National Astronomical Observatories, Chinese Academy of Sciences, Beijing 100101, China
\and School of Astronomy, Institute for Research in Fundamental Sciences (IPM), P. O. Box 19395-5531, Tehran, Iran
\and School of Astronomy and Space Science, University of Chinese Academy of Sciences, Beijing 100049, China
}
\date{Received: date / Revised version: date}
%
\abstract{
QPOs are seen as peak features in the X-ray power spectral density of stellar mass black holes and neutron stars, of which frequencies can be measured with high precision. These QPO frequencies are believed to be related to fundamental frequencies of test particles, which are mainly determined by the background metric. We consider the metric introduced in [Ghasemi-Nodehi and Bambi, Eur. Phys. J. C 76, 290 (2016)]. The fundamental frequencies in this metric are sensitive to some of the metric parameters but insensitive to other parameters, which means, the differences in fundamental frequencies in this metric and those in the Kerr ones can be significant for small changes of some but not all parameters around the Kerr value. By comparing with the QPO observations of GRO J1655-40, we find that only one parameter of the Ghasemi-Nodehi-Bambi metric can be strongly constrained, but other parameters cannot. We also use nested algorithm to investigate whether better constraints on the Ghasemi-Nodehi- Bambi metric parameters can be obtained from QPO observations of multiple objects by simulations. We find that four parameters can be strongly constrained while other parameters cannot . Our results suggest that QPOs may be important tools for testing the Kerr metric.%
\PACS{
      {PACS-key}{discribing text of that key}   \and
      {PACS-key}{discribing text of that key}
     } 
} 
\authorrunning{M. Ghasemi-Nodehi, et al.}
\maketitle
%


\section{Introduction}

Einstein theory of General Relativity (GR) was born more than 100 years ago and successfully passed test in the weak gravity field regime \cite{einstein1916,t1,t2,t3,t4}. According to GR, astrophysical black holes (BHs) are described by the Kerr metric with only two parameters, i.e., mass and spin \cite{k1,k2}. However, the nature of the Kerr BH is still to be verified.  Important test of GR in the strong gravity field regime can be from direct observational confirmation of the Kerr metric, via either
electromagnetic wave \cite{e1,e2} or gravitational wave \cite{g1,g2} observations. The electromagnetic tests in the literature mainly make use of the iron line method \cite{ka} and/or continuum fitting method \cite{cfm} by utilizing X-ray data \cite{r1,r2}. Black hole shadow also offers a unique tool to probe the event horizon and nature of BHs, which has been demonstrated by recent observations of M87 through the Event Horizon Telescope (EHT) \cite{evh,evh1}.

There have also been a number of attempts to generalize the Kerr solution by phenomenological parameterization \cite{p1,p2,p3,p4,p5,p6,p7,p8,p9,p10,p11,p12,p13,p14,p15,p16}. The idea is mostly to write a more general metric that includes the Kerr one as its special case, similar to the PPN of the Schwarzschild metric \cite{t1}.
Generally, parameters of the non-Kerr metric may be degenerated in testing strong gravity fields. Therefore, the Kerr metric cannot be verified only by the detection of its characteristic properties, and it would be proofed only when all deformation parameters of the non-Kerr metric are required to vanish by observations.

Quasi-periodic oscillations (QPOs), peak features in the X-ray power spectral density of stellar mass BHs and neutron stars, are recently recognized as promising tools to test GR in the strong gravity field regime \cite{q1,q2,q3,q4,q5,q6,q7}. QPOs have been detected in four systems GRO J1655-40, XTE J1550-564, GRS 1915+105, and H1743-322, they are not sensitive to the properties of accretion flows. The exact physical mechanism responsible for the production of the QPOs is not clear, though a number of  different scenarios have been proposed, including relativistic precession model \cite{rpm}, diskoseismology models \cite{dm}, resonance models \cite{rm}, and p-mode oscillations of an accretion torus \cite{rezz}. In most scenarios, QPO frequencies are related to the characteristic orbital frequencies of test particles, which are determined only by the BH metric. The main advantage of using QPOs with respect to other methods is that the QPO frequencies can be measured with high accuracy. Thus, QPOs can provide a unique tool to probe the spacetime and geometry around the central compact objects.

One of us proposed a new parameterization to Kerr metric in Reference \cite{p15} (hereafter the GB metric). The Kerr case would be recovered by setting all the parameters involved in the GB metric equal to $1$. Any parameter different from $1$ indicates a deviation from the Kerr case. Assuming the GB metric, we have studied the BH shadow in the same paper and X-ray reflection spectroscopy in  \cite{ghasemi}. BH shadow studies show that only $4$ parameters involved in the GB metric 
have weak impact on the shadow shape while other parameters have almost no impact, and thus it will be difficult to constrain the GB metric by using BH shadow measurements. In the paper \cite{ghasemi}, one of us studied the X-ray reflection spectroscopy resulting from accretion systems to constrain the parameters in the GB metric. Our results show that all parameters except one can be potentially well constrained with near future X-ray missions. In the present paper, we investigate the impact of GB parameters of Ref. \cite{p15} on QPOs. Our results show significant parameter degeneracy in constraining the GB parameters.

The content of this paper is as follows. In Section~\ref{theory}, we provide the theoretical framework of our calculations. Results and discussions are written in Section~\ref{res}. Summary and conclusions are presented in section~\ref{summary}. Throughout the paper, we use units in which $G_N = c = 1$.


\section{Theoretical Framework  \label{theory}}

Three fundamental frequencies characterize the equatorial orbit of a test particle rotating around a BH, i.e., the Keplerian frequency $\nu_{\phi}$, radial epicyclic frequency $\nu_r$, and vertical epicyclic frequency  $\nu_{\theta}$. These frequencies only depend on the BH metric and radius of orbit.

The general case can be studied by the line element
\be
ds^2 = g_{tt} dt^2 + g_{rr}dr^2 + g_{\theta\theta} d\theta^2
+ 2g_{t\phi}dt d\phi + g_{\phi\phi}d\phi^2\,, \nonumber \\
\ee
that is for a generic stationary and axisymmetric spacetime and the metric components are independent of $t$ and $\phi$.

There are two conserved quantities associated to $t$ and $\phi$, specific energy at infinity, $E$, and z-component of specific angular momentum at infinity, $L_z$. The $t$ and $\phi$ components of 4-velocity of a test particle are~\cite{bookbambi}
\be
\dot{t} &=& \frac{E g_{\phi\phi} + L_z g_{t\phi}}{
g_{t\phi}^2 - g_{tt} g_{\phi\phi}}  \\
\dot{\phi} &=& - \frac{E g_{t\phi} + L_z g_{tt}}{
g_{t\phi}^2 - g_{tt} g_{\phi\phi}} \, .
\ee
One may introduce $V_{\rm eff}$ from the conservation of the rest mass $g_{\mu\nu}\dot{x}^\mu \dot{x}^\nu = -1$,
\be
\label{geod}
g_{rr}\dot{r}^2 + g_{\theta\theta}\dot{\theta}^2
= V_{\rm eff}(r,\theta,E,L_z) \, ,
\ee
where
\be
V_{\rm eff} = \frac{E^2 g_{\phi\phi} + 2 E L_z g_{t\phi} + L^2_z
g_{tt}}{g_{t\phi}^2 - g_{tt} g_{\phi\phi}} - 1  \, .
\ee

Geodesic equation can also be written as
\be
\frac{d}{d\lambda} \left( g_{\mu\nu} \dot{x}^\nu \right)
= \frac{1}{2} \left( \partial_\mu g_{\nu\rho} \right) \dot{x}^\nu \dot{x}^\rho \, .
\ee
Circular orbit requires that $\dot{r} = \dot{\theta} = \ddot{r} = 0$, so the radial component is
\be
\label{eq-geo}
\left( \partial_r g_{tt} \right) \dot{t}^2
+ 2 \left( \partial_r g_{t\phi} \right) \dot{t} \dot{\phi}
+ \left( \partial_r g_{\phi\phi} \right) \dot{\phi}^2 = 0 \, .
\ee
This equation gives us the orbital angular velocity
\be
\Omega_\phi  = \frac{\dot{\phi}}{\dot{t}} = \frac{- \partial_r g_{t\phi}
\pm \sqrt{\left(\partial_r g_{t\phi}\right)^2
- \left(\partial_r g_{tt}\right) \left(\partial_r
g_{\phi\phi}\right)}}{\partial_r g_{\phi\phi}} \, ,
\ee
where the sign is ($+/-$) for corotating/counterrotating  orbits and orbital frequency can be calculated as $\nu_\phi= \Omega_\phi/2\pi$.

Considering $\dot{r} = \dot{\theta} = 0$ in $g_{\mu\nu}\dot{x}^\mu \dot{x}^\nu = -1$, we have
\be
\dot{t} = \frac{1}{\sqrt{-g_{tt} - 2g_{t\phi}\Omega_\phi - g_{\phi\phi}\Omega^2_\phi}} \, .
\ee
We find $E$ and $L_z$ since $-E = g_{tt} \dot{t} + g_{t\phi} \dot{\phi}$ and $L_z = g_{t\phi} \dot{t} + g_{\phi\phi} \dot{\phi}$,
\be
E &=& - \frac{g_{tt} + g_{t\phi}\Omega_\phi}{
\sqrt{-g_{tt} - 2g_{t\phi}\Omega_\phi - g_{\phi\phi}\Omega^2_\phi}} \, , \\
L_z &=& \frac{g_{t\phi} + g_{\phi\phi}\Omega_\phi}{
\sqrt{-g_{tt} - 2g_{t\phi}\Omega_\phi - g_{\phi\phi}\Omega^2_\phi}} \, .
\ee
Substituting $E$ and $L_z$ in equation~(\ref{geod}) together with metric component gives us the $V_{\rm eff}$.

From equation~(\ref{geod}) we consider small perturbation around circular orbit along radial and vertical directions at linear order. If $\delta_r$ and $\delta_\theta$ are the small displacements around the mean orbit, $r = r_0 + \delta_r$ and $\theta = \pi/2 + \delta_\theta$, and neglecting terms $O(\delta_r^2)$ and $O(\delta_\theta^2)$, we have following differential equations
\be
\label{eq-o1}
&&\frac{d^2 \delta_r}{dt^2} + \Omega_r^2 \delta_r = 0,  \\
&&\frac{d^2 \delta_\theta}{dt^2} + \Omega_\theta^2 \delta_\theta = 0 \, ,
\label{eq-o2}
\ee
where
\be
\label{eq-or}
&&\Omega^2_r = - \frac{1}{2 g_{rr} \dot{t}^2}
\frac{\partial^2 V_{\rm eff}}{\partial r^2},  \\
&&\Omega^2_\theta = - \frac{1}{2 g_{\theta\theta} \dot{t}^2}
\frac{\partial^2 V_{\rm eff}}{\partial \theta^2} \, .
\label{eq-ot}
\ee
The radial epicyclic frequency and the vertical epicyclic frequency are, respectively, $\nu_r = \Omega_r/2\pi$ and $\nu_\theta = \Omega_\theta/2\pi$.

In this paper, we consider relativistic precession model that recent studies seem to support it \cite{Motta1,Motta2}.
This model first proposed to explain the frequency of the QPOs in power spectral density of neutron star and then is extended to the QPOs of the stellar mass BH \cite{rpm}. This model simply relates the observed frequencies with the orbital motion of a test particle. One can find the periastron precession frequency, $\nu_p$, and nodal precession frequency from fundamental frequencies,
\be
\nu_p &=& \nu_\phi - \nu_r\,,  \\
\nu_n &=& \nu_\phi - \nu_{\theta}\, .
\ee
These frequencies can be measured observationally and compared with the fundamental frequencies computed from the background metric. This provides a unique opportunity with high accuracy to probe the strong gravity regime and test possible deviations from the
Kerr solution of GR.

In Ref~\cite{p15} one of us introduced 11 parameters, $b_i$ as follows,
\be
\label{eq-m}
ds^2 &=& - \left( 1 - \frac{2 b_1 M r}{r^2 + b_2 a^2 \cos^2\theta} \right) dt^2 \nonumber\\
&&
- \frac{4 b_3 M a r \sin^2\theta}{r^2 + b_4 a^2 \cos^2\theta} dt d\phi
+ \frac{r^2 + b_5 a^2 \cos^2\theta}{r^2 - 2 b_6 M r + b_7 a^2} dr^2
\nonumber\\
&&
+ \left( r^2 + b_8 a^2 \cos^2\theta \right) d\theta^2
\nonumber\\
&&
+ \left( r^2 + b_9 a^2 + \frac{2 b_{10} M a^2 r
\sin^2\theta}{r^2 + b_{11} a^2 \cos^2\theta} \right) \sin^2\theta d\phi^2 \, .
\ee
Kerr case recovers for all $b_i = 1$. Any differences from $b_i = 1$ show deviations from general relativity. $b_1$ can be set as one because it is coefficient of mass, $b_1 M$ and also $b_3 = 1$ in the same way asymptotic specific angular momentum is $b_3 a$. We also have $b_6$ close to one from solar system experiment. So we set $b_1 = b_3 = b_6 = 1$ and we do not consider these three parameters in our QPO calculations.


\section{Results}\label{res}

QPOs showing in the X-ray power spectral density of stellar mass BHs and neutron stars provide precise information about the background metric. QPOs are frequencies measured for a BH system are believed to be determined by the BH metric but not the accretion flow, though it may model dependent. In this paper, we adopt the relativistic precession model and assume that QPO frequencies are related to the characteristic orbital frequencies of test particles. These characteristic frequencies are solely determined by the BH metric. We also assume that the GB metric describes the BH metric, for which the Kerr metric is a special case.

\begin{figure*}
\vspace{0.4cm}
\begin{center}
\includegraphics[type=pdf,ext=.pdf,read=.pdf,width=8.0cm]{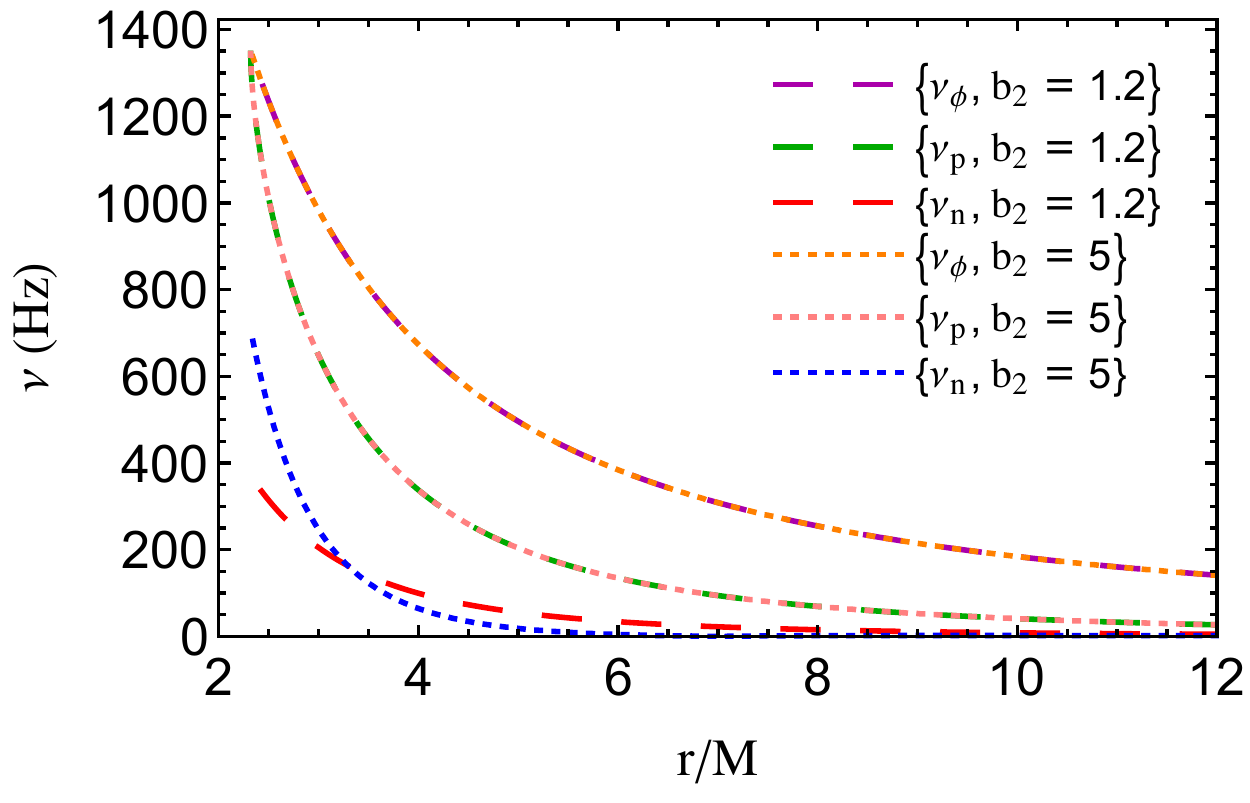}
\hspace{0.8cm}
\includegraphics[type=pdf,ext=.pdf,read=.pdf,width=8.0cm]{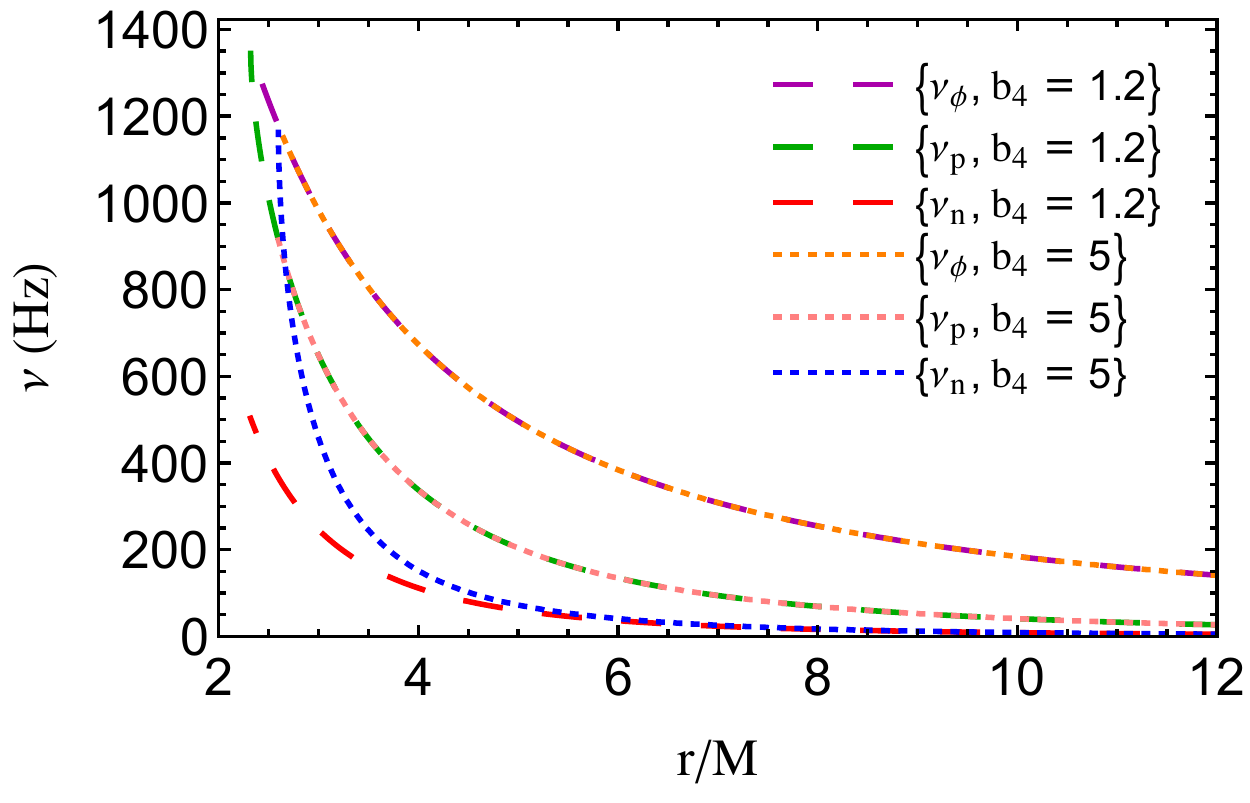}
\vspace{0.8cm}
\includegraphics[type=pdf,ext=.pdf,read=.pdf,width=8.0cm]{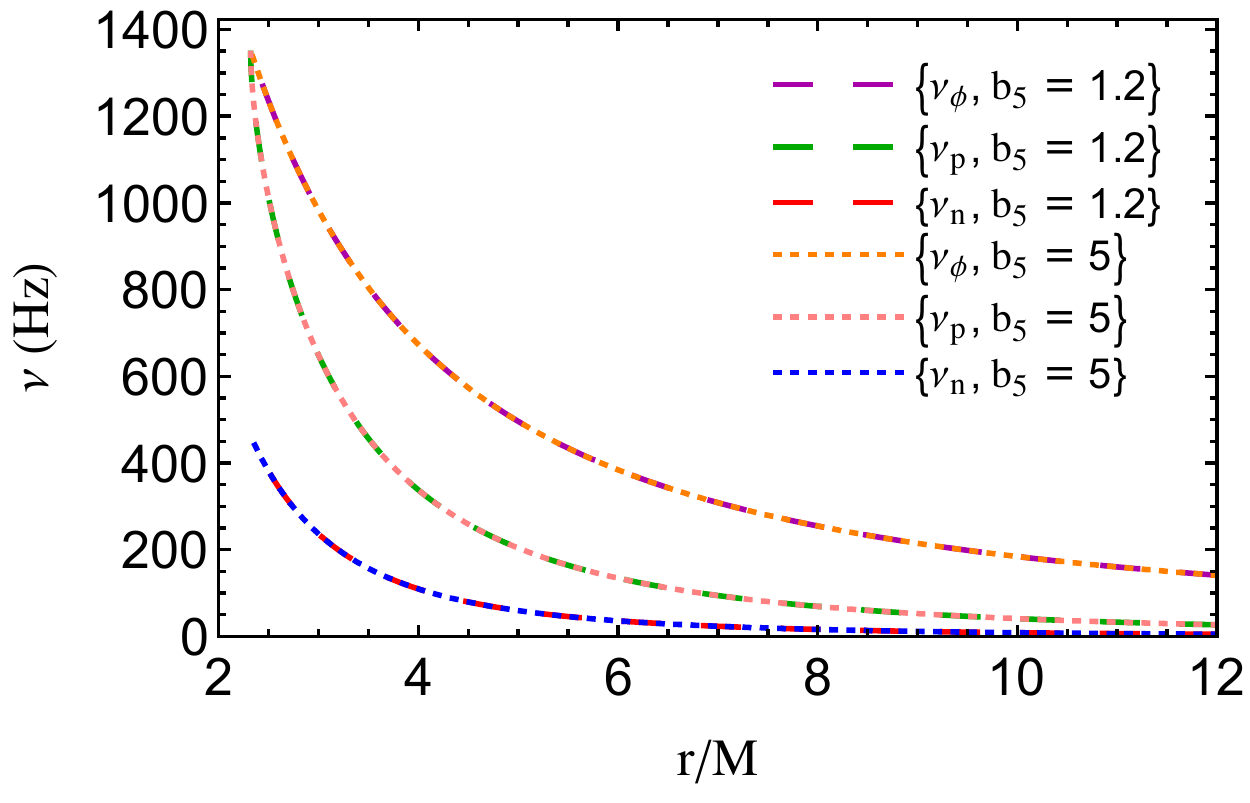}
\hspace{0.8cm}
\includegraphics[type=pdf,ext=.pdf,read=.pdf,width=8.0cm]{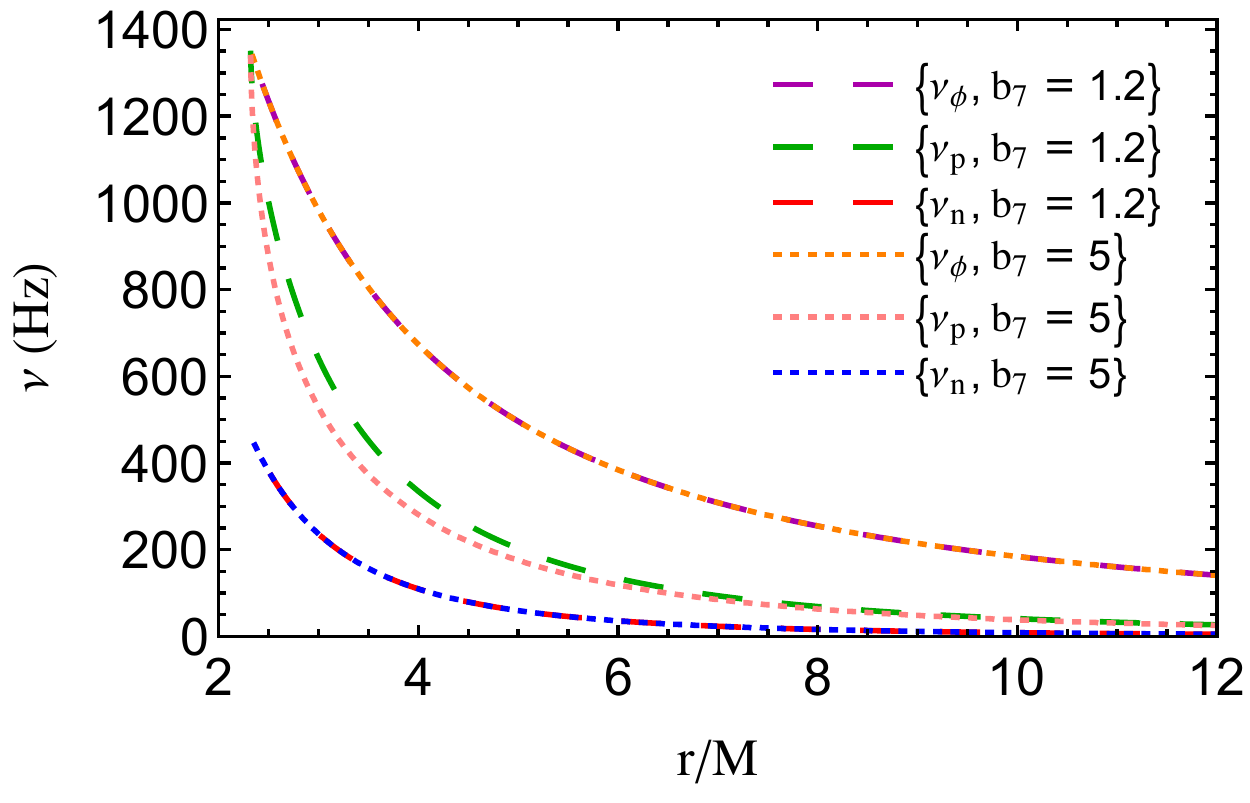}
\end{center}
\vspace{-0.3cm}
\caption{Fundamental frequencies of a test particle on a circular orbit at different radii in the GB metric. Top-left, top-right, bottom-left, and bottom right panels show those frequencies obtained for a GB BH with $(a/M, b_2)= (0.9, 1.2)$ and $ (0.9, 5)$, $(a/M, b_4)= (0.9, 1.2)$ and $ (0.9, 5)$, $(a/M, b_5)= (0.9, 1.2)$ and $(0.9, 5)$, and $(a/M, b_7)= (0.9, 1.2)$ and $(0.9, 5)$, respectively. All other not mentioned $b_i$ in each case equal to $1$. In each panel, purple long-dashed, green short-dotted, and red dotted lines show $\nu_\phi$, $\nu_p$, and $\nu_n$ for case $b_i = 1.2$, respectively. The purple, green, and red solid lines show $\nu_\phi$, $\nu_p$, and $\nu_n$ for case $b_i = 1.2$, respectively. The mass and dimensionless spin parameter of the black hole are fixed at $5.4M_\odot$ and $0.9$ .
\label{freq1}}
\end{figure*}

\begin{figure*}
\vspace{0.4cm}
\begin{center}
\includegraphics[type=pdf,ext=.pdf,read=.pdf,width=8.0cm]{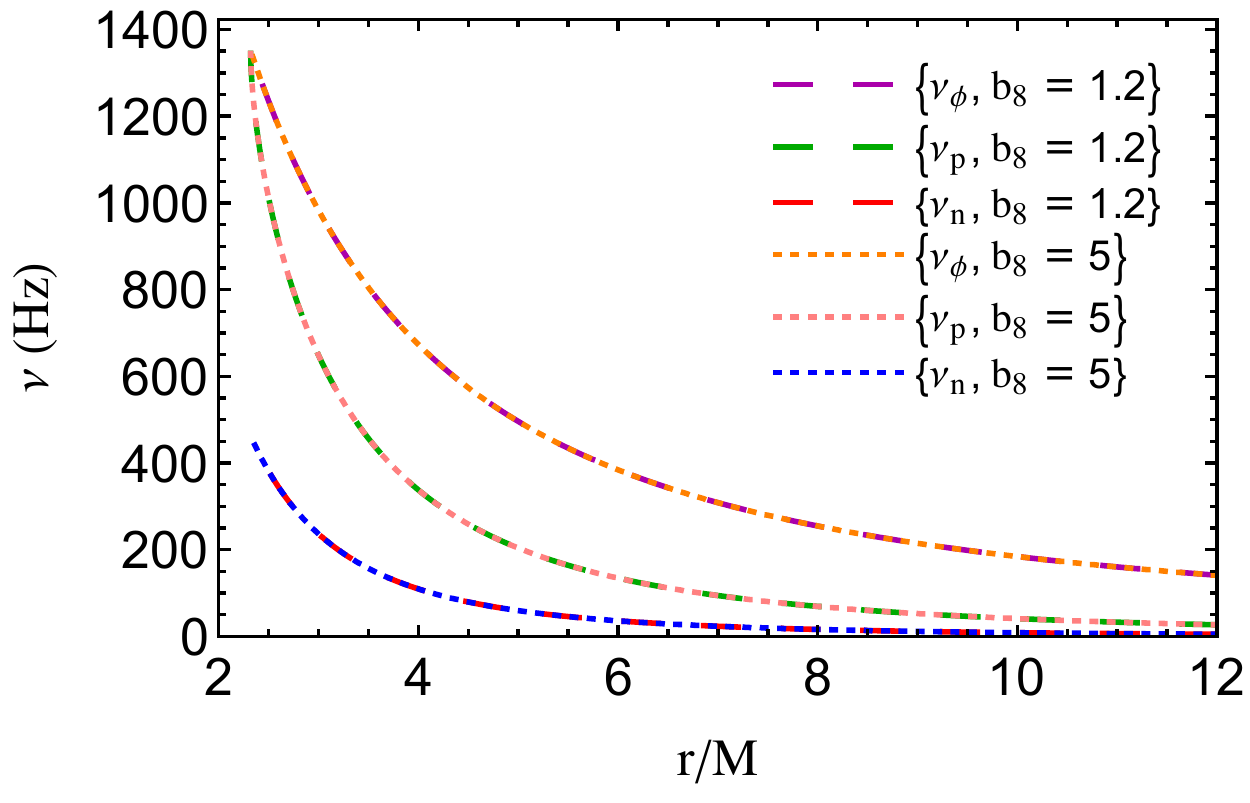}
\hspace{0.8cm}
\includegraphics[type=pdf,ext=.pdf,read=.pdf,width=8.0cm]{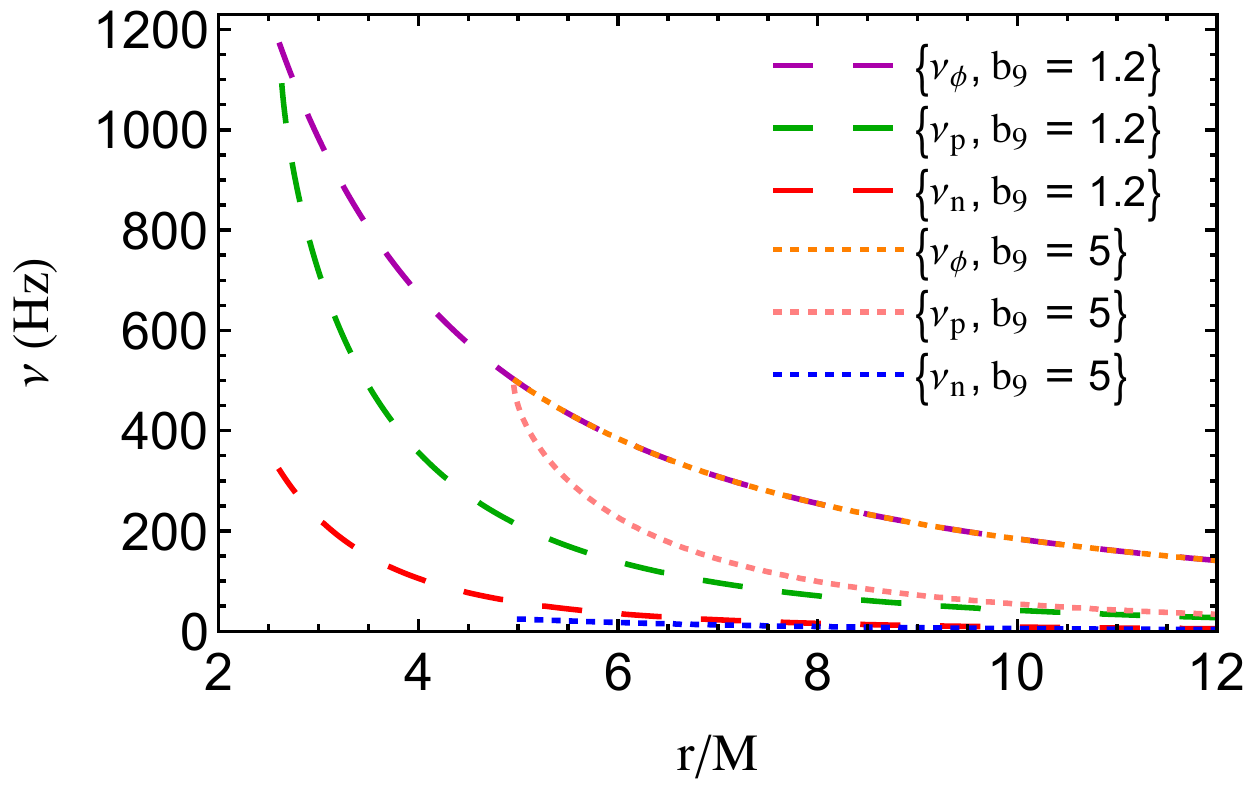}
\hspace{0.8cm}
\includegraphics[type=pdf,ext=.pdf,read=.pdf,width=8.0cm]{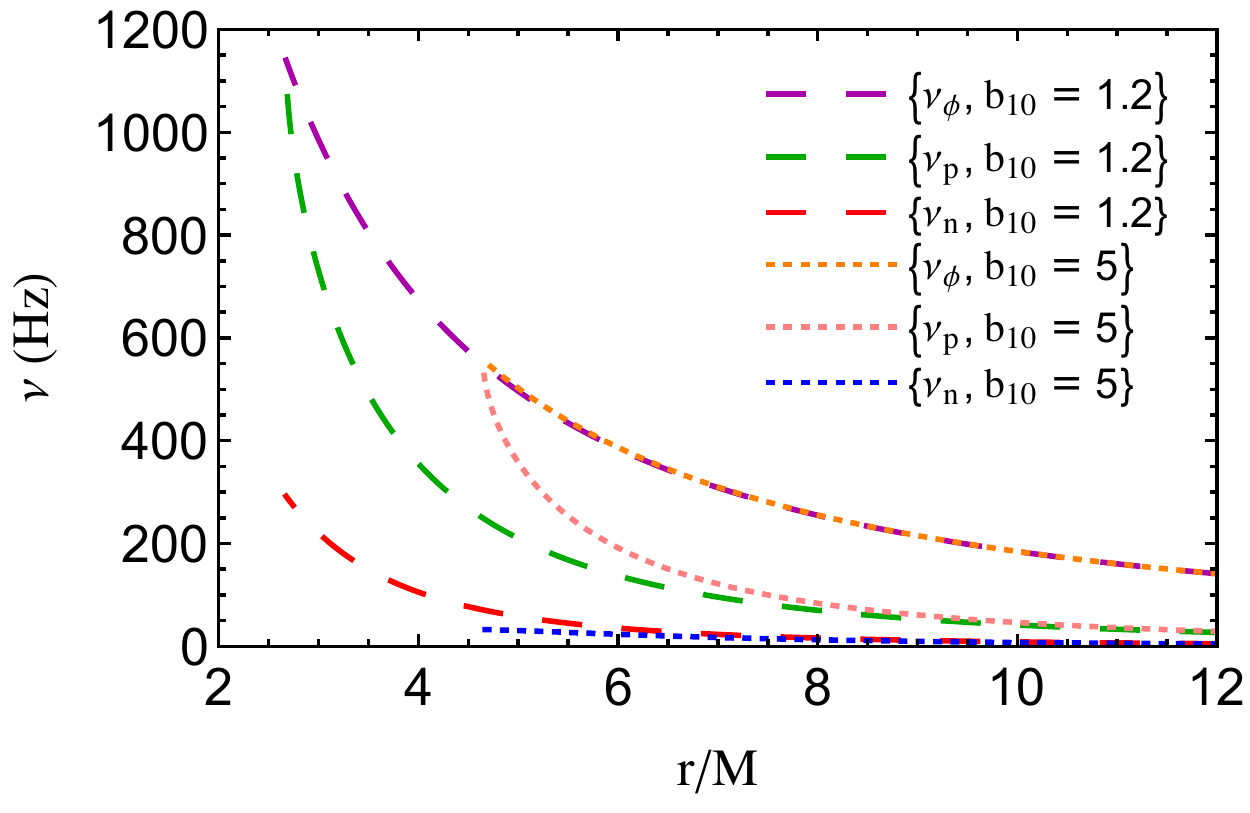}
\hspace{0.8cm}
\includegraphics[type=pdf,ext=.pdf,read=.pdf,width=8.0cm]{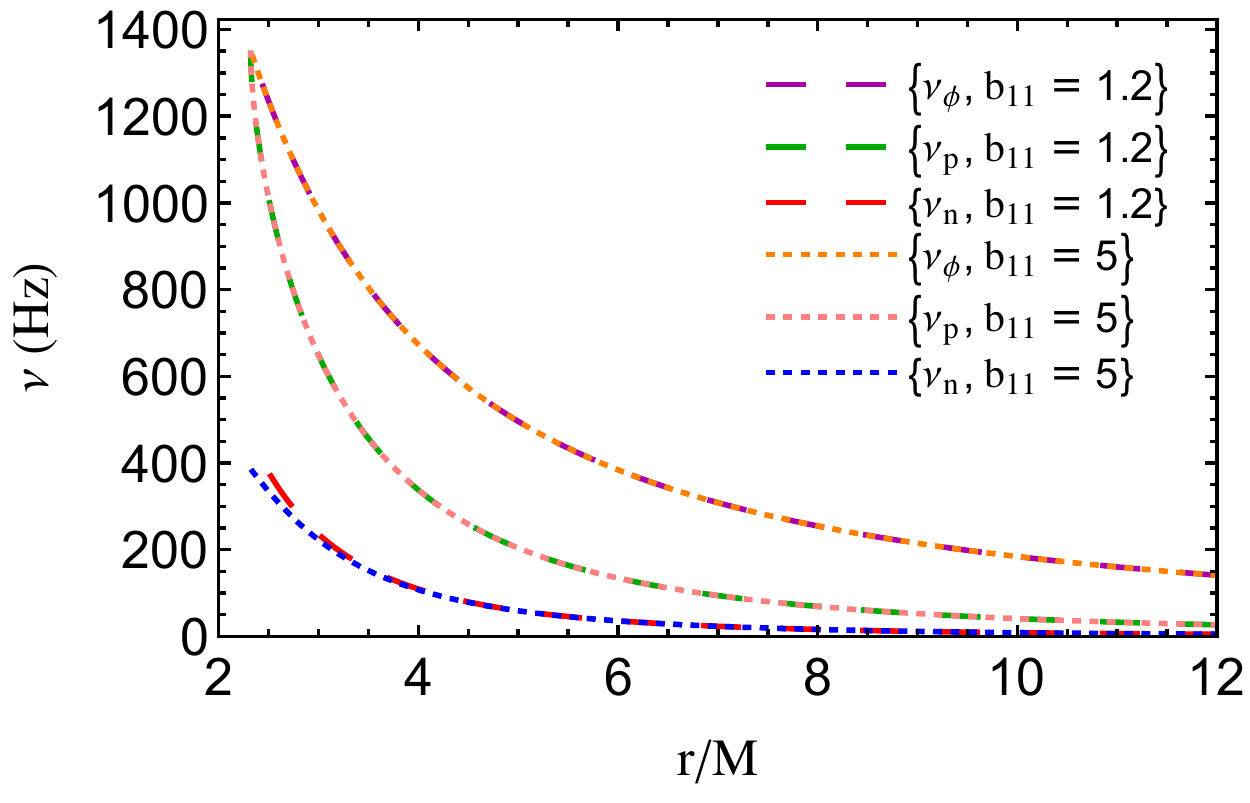}
\end{center}
\vspace{-0.3cm}
\caption{Legend similar to Figure~\ref{freq2}, while top-left, top-right, bottom-left, and bottom right panels show those frequencies obtained for a GB BH with $(a/M, b_8)= (0.9, 1.2)$ and $(0.9, 5)$, $(a/M, b_9)= (0.9, 1.2) $ and $ (0.9, 5)$, $(a/M, b_{10})= (0.9, 1.2)$ and $(0.9, 5)$, and $(a/M, b_{11})= (0.9, 1.2) $ and $ (0.9, 5)$, respectively.
\label{freq2}}
\end{figure*}

We first calculate frequencies $\nu_\phi$, $\nu_r$, $\nu_\theta$, $\nu_p$, and $\nu_n$ for the GB metric with different settings of the metric parameters.
Figures~\ref{freq1} and \ref{freq2} show $\nu_\phi$, $\nu_p$, and $\nu_n$ as a function of radius $r$ for GB BHs with mass $5.4M_\odot$, dimensionless spin parameter $a_*= a/M = 0.9$, but different $b_i$ (not too much different from the Kerr case $b_i=1$). As seen from these two Figures, different settings of $b_2$ or $b_4$ (e.g., $b_2$ or $b_4=1.2$ and $5$) may lead to substantial difference in the resulting $\nu_n$ but negligible differences in $\nu_\phi$ and $\nu_p$ (top left and right panels of Fig.~\ref{freq1}); different settings of $b_7$, $b_9$, or $b_{10}$ (e.g., $1.2$ and $5$) may lead to significant difference in the resulting $\nu_p$ but not $\nu_\phi$ and $\nu_n$ (bottom-right panel of Fig.~\ref{freq1}, top-right or bottom-left panel of Fig.~\ref{freq2}); different settings of $b_5$, $b_8$, or $b_{11}$ may only lead to little differences in all resulting frequencies (bottom-left panel of Fig.~\ref{freq1}, top-left or bottom-right panel of Fig.~\ref{freq2}).
We emphasize here that different settings of $b_i$ may lead to significantly different innermost stable circular orbits (ISCOs) and thus the frequency curves cut at different $r/M$, which are most significant in the top-right and bottom-left panels of Figure~\ref{freq2} for $b_9$ and $b_{10}$, respectively. Such cutoffs are crucial in obtaining constraints on $b_i$ if QPOs are originated from the inner edge of accretion disks.

In the Kerr metric, the three fundamental frequencies can be derived analytically as,
\be
\label{kerrfre}
\nu_{\phi} &=& \left(\frac{1}{2\pi} \right) \frac{M^{1/2}}{r^{3/2}\pm aM^{1/2}}\,, \\
\label{kerrfre1}
\nu_r &=& \nu_{\phi} \left( 1- \frac{6M}{r} \pm \frac{8aM^{1/2}}{r^3/2} - \frac{3a^2}{r^2} \right)^{1/2}\,,  \\
\label{kerrfre2}
\nu_{\theta} &=& \nu_{\phi} \left( 1 \mp \frac{4aM^{1/2}}{r^{3/2}}+\frac{3a^2}{r^2} \right)^{1/2}\,.
\ee
The Schwarzschild case would recover by imposing $a = 0$, i.e.,
\be
\nu_{\phi} &=& \nu_{\theta} = \frac{1}{2\pi} \frac{M^{1/2}}{r^{3/2}}\,,  \\
\nu_r &=& \nu_{\phi} \left( 1- \frac{6M}{r} \right)^{1/2}\,.
\ee

\begin{figure*}
\vspace{0.4cm}
\begin{center}
\includegraphics[type=pdf,ext=.pdf,read=.pdf,width=8.0cm]{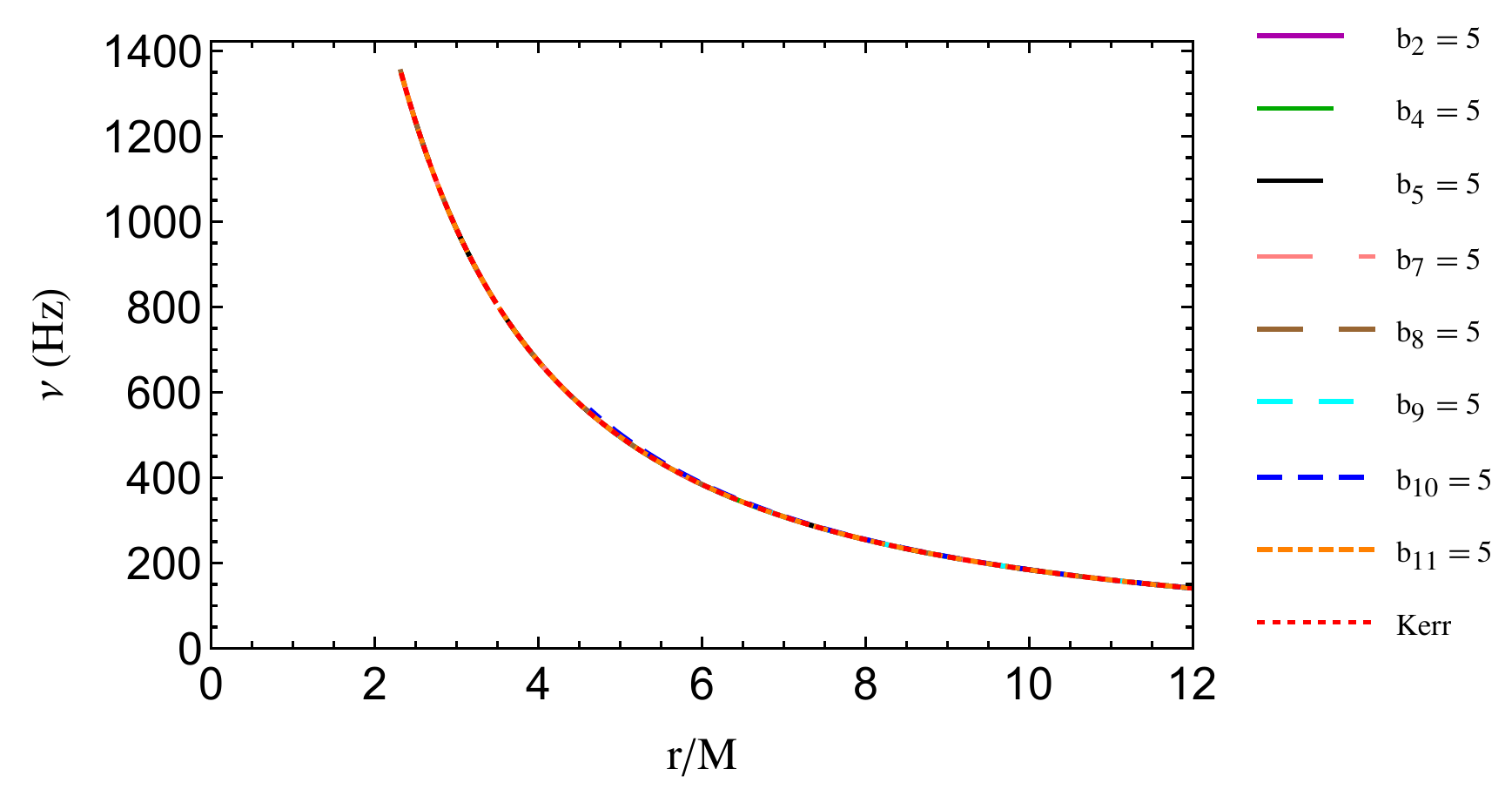}
\hspace{0.8cm}
\includegraphics[type=pdf,ext=.pdf,read=.pdf,width=8.0cm]{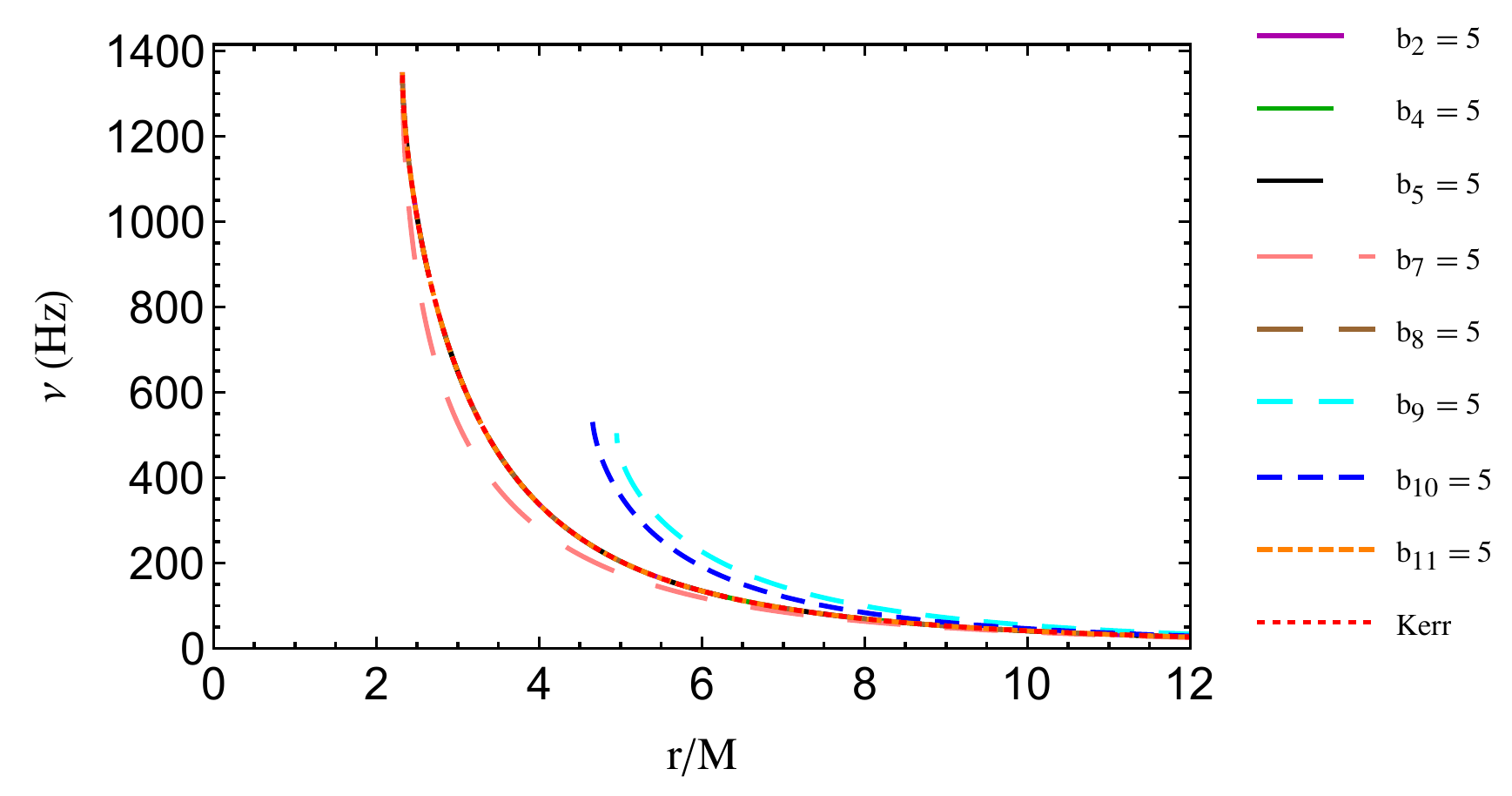}\\
\hspace{0.8cm}
\includegraphics[type=pdf,ext=.pdf,read=.pdf,width=8.0cm]{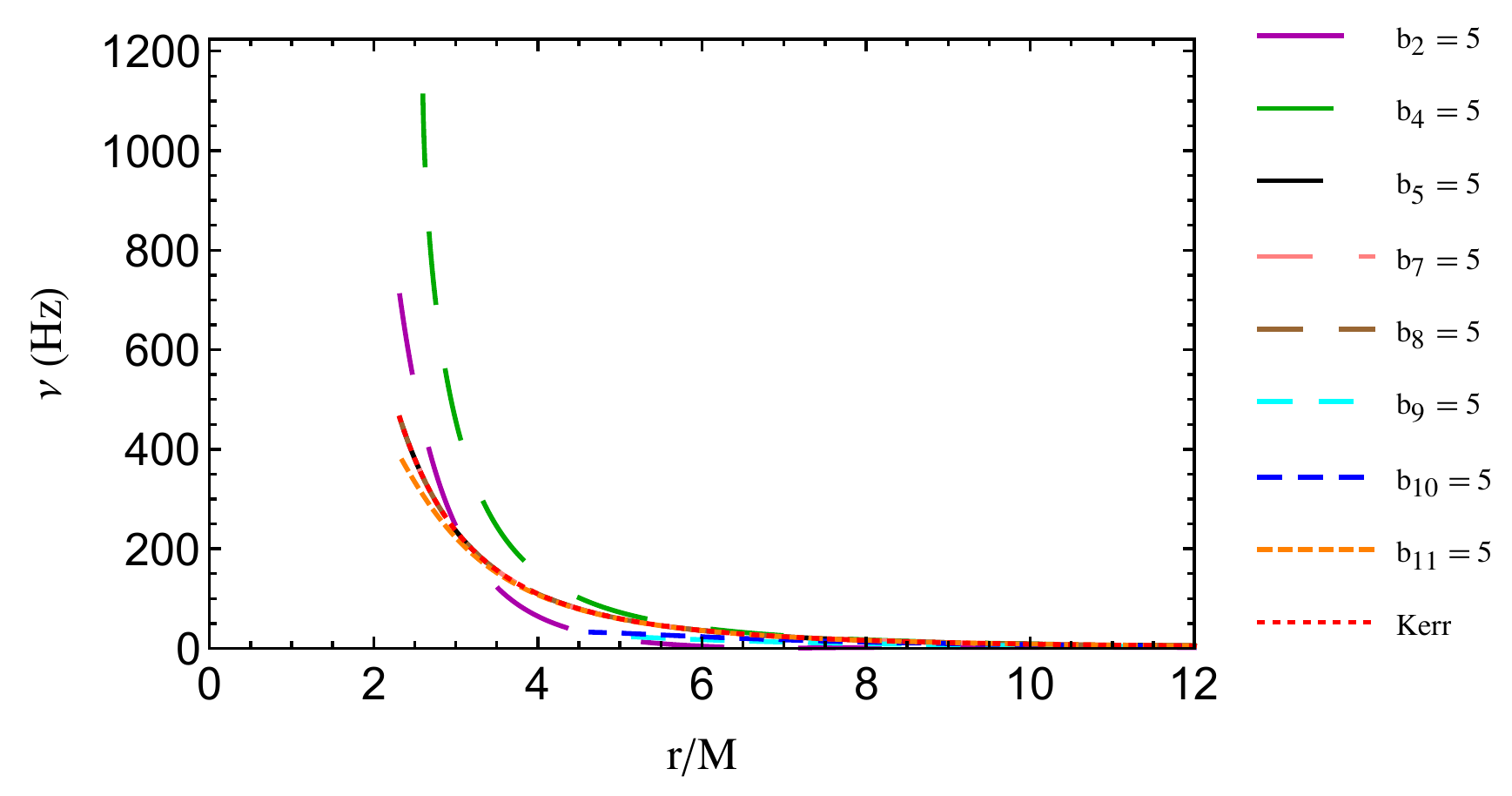}
\vspace{-0.3cm}
\caption{Impact of all GB parameters on the shape of $\nu_\phi$ (top left panel), $\nu_p$ (top right panel), $\nu_n$ (bottom panel). The spin parameter is $0.9$ and mass is considered as $5.4 M_\odot$. See the text for more details.
\label{freq3}}
\end{center}
\end{figure*}

In the Kerr metric $\nu_{\theta} \geq \nu_r$, and in the Schwarzschild metric $\nu_{\phi} = \nu_{\theta} > \nu_r$. In our calculation for $b_i = 1.2$, similar to the Kerr case and Schwarzschild case all $\nu_\theta$s are greater than $\nu_r$, but for the case $b_2 = 5$,  $\nu_r \geq \nu_\theta$. This is possible, for example, if the ISCO is marginally vertically stable, $\nu_\theta = 0$ at the ISCO. Thus, we have $\nu_r > \nu_{\theta}$. In order to show differences with respect to the Kerr case, we plot Figure~\ref{freq3}. This figure shows the impact of different GB parameters on the shape of $\nu_\phi$ (top left panel), $\nu_p$ (top right panel), $\nu_n$ (bottom panel). For all cases, the mass and dimensionless spin parameter are set to $5.4 \,M_\odot$ and $0.9$, respectively. It is obvious that $\nu_\phi$ curves for GB metric with $b_i=5$ ($i=2, 4, 5, 7, 8, 9, 10$, and $11$, respectively) are almost the same as that for the Kerr case with all $b_i=1$ (top-left panel), while $\nu_p$ curves for the cases with $b_7= 5$, $b_9=5$, or $b_{10}=5$ are significantly different from those for other cases (almost identical; top-right panel), and $\nu_n$ curves for the cases with $b_2= 5$, $b_4= 5$, $b_9=5$, or $b_{10}=5$ are significantly different from those for other cases (almost identical; bottom panel).

According to Figures~\ref{freq1}, \ref{freq2} and \ref{freq3}, we conclude that the GB metric can be distinguished from the Kerr one if any of the GB parameters $b_2$, $b_9$, $b_{10}$ is substantially different from $1$. The reason is that some of the resulting fundamental frequencies from such a GB metric are substantial different from those of the Kerr one. The GB parameters $b_4$ and $b_7$ may also be constrained but with less significance since the differences in the fundamental frequencies are smaller. 

We note here that $b_{11}$ may also be constrained since the QPO frequencies may be only significantly affected by $b_{11}$ when BH spin is extremely high, which is discussed later in Section~\ref{subsec:mock} and can be seen from Figure~\ref{b_11-a}.

Other GB parameters like $b_5$ and $b_8$ cannot be constrained because of the negligible differences in fundamental frequencies for different settings of those $b_i$.

\subsection{QPO observations for GRO J1655-40}

Rossi X-ray Timing Explorer (RXTE) mission observations of GRO J1655-40 have revealed that it has two high frequency QPOs and one low frequency QPO \cite{Motta1,gro2}. These frequencies are measured with high accuracies, i.e.,
$\nu_{1} = 441^{+2}_{-2}$\,Hz, $\nu_2 = 298^{+4}_{-4}$\,Hz, and $\nu_3 = 17.3^{+0.1}_{-0.1}$\,Hz.
If we assume that these frequencies correspond to $\nu_\phi$, $\nu_p$, and $\nu_n$, respectively, and the central BH is described by the Kerr metric, these frequencies should be determined by three unknown parameters $M$, $a$, and $r$ as seen from Equations~(\ref{kerrfre})-(\ref{kerrfre2}). Therefore, strong constraints should be able to obtain on the mass $M$ and dimensionless spin $a$ of the BH.

If the BH metric in GRO J1655-40 is described by the GB metric rather than the Kerr metric, the determined QPO frequencies may be used to put constraint on the GB parameters $b_i$s. For simplicity, we assume that only one of the $b_i$s is different from that of the Kerr case, i.e., $b_i\neq 1$ and $b_j$=1 with $j\neq i$, in order to get a meaningful constraint. For each given $i$ ($2$, $4$, $5$, $7$, $8$, $9$, $10$, or $11$), we calculate $\chi^2$ as follows to obtain a constraint on the $b_i$
\be
\chi^2 (a_*, b_i) &=& \left( \frac{\nu_{\phi}-441}{2} \right)^2 + \left( \frac{\nu_{p}-298}{4} \right)^2 \nonumber\\
&&  + \left( \frac{\nu_{n}-17.3}{0.1}\right)^2 \,,
\label{chi2}
\ee
where $\nu_\phi, \nu_n$, and $\nu_p$ are, respectively, the orbital, nodal, and periastron frequency. The numbers $441$, $298$, and $17.3$ are the QPO frequencies measured from the X-ray observations of GRO J1655-40, and $2$, $4$, and $0.1$ are the corresponding $1$-$\sigma$ errors $\sigma_i$ here. We set $M = 5.4\,M_{\odot}$ here, which is obtained from independent optical observations \cite{opt}.  Here we adopt the $\chi^2$ statistics to obtain such constraints.

\begin{figure*}
\vspace{0.4cm}
\begin{center}
\includegraphics[type=pdf,ext=.pdf,read=.pdf,width=8.0cm]{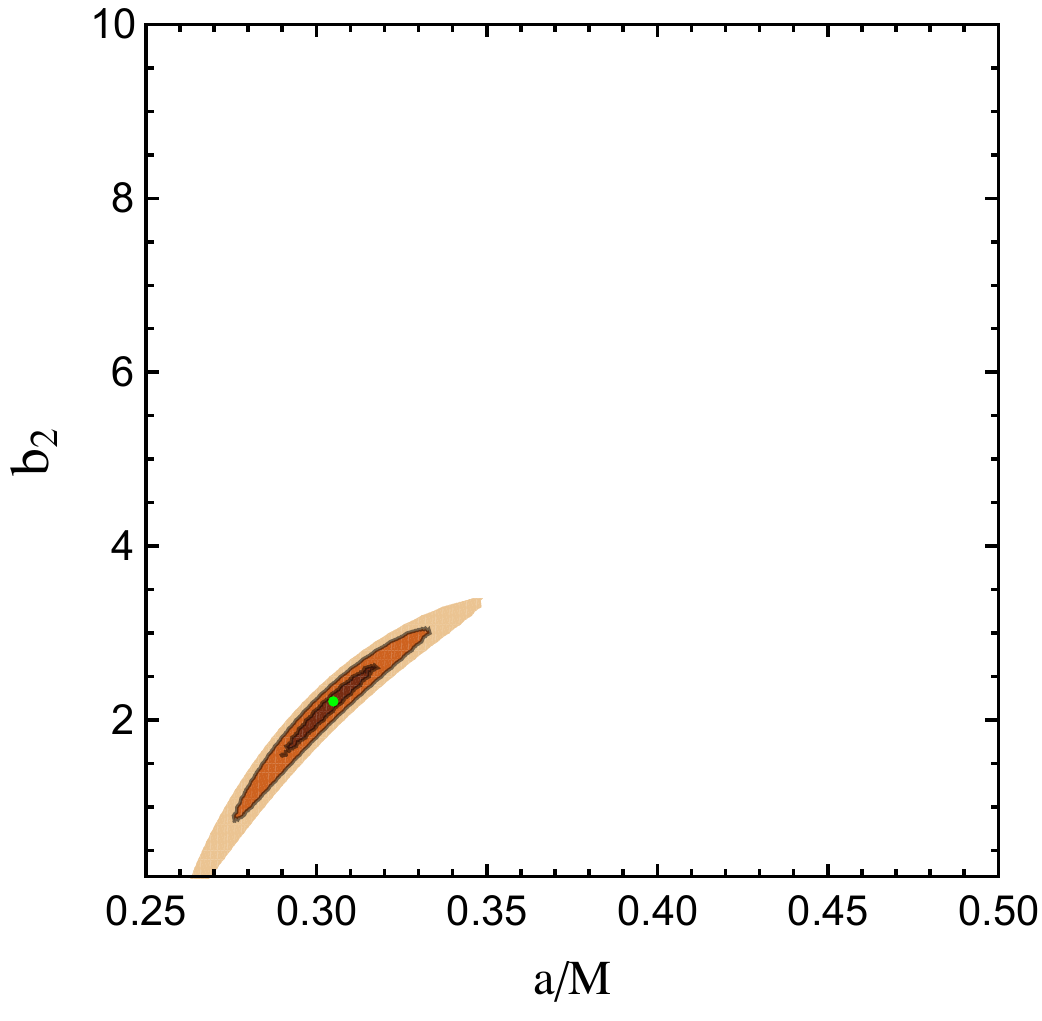}
\hspace{0.8cm}
\includegraphics[type=pdf,ext=.pdf,read=.pdf,width=8.0cm]{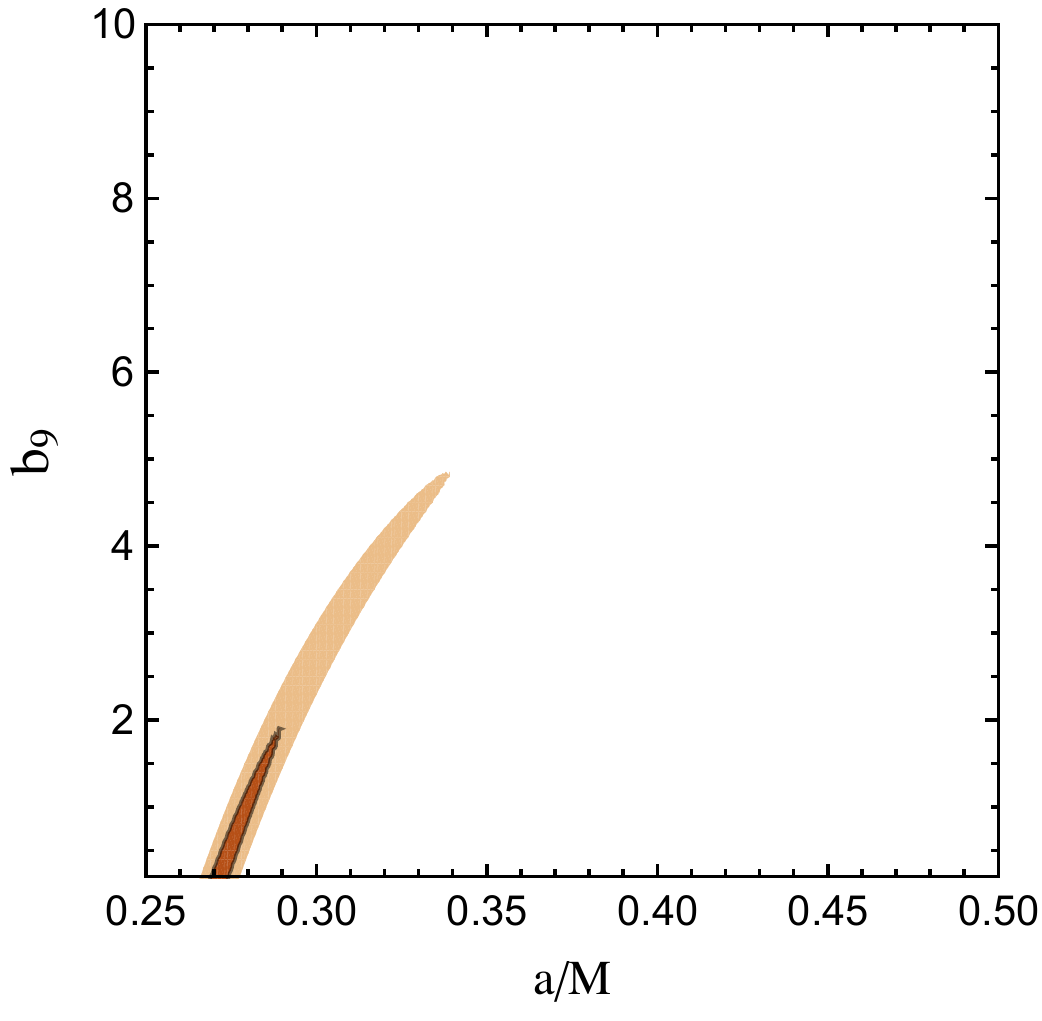}\\
\vspace{0.8cm}
\includegraphics[type=pdf,ext=.pdf,read=.pdf,width=8.0cm]{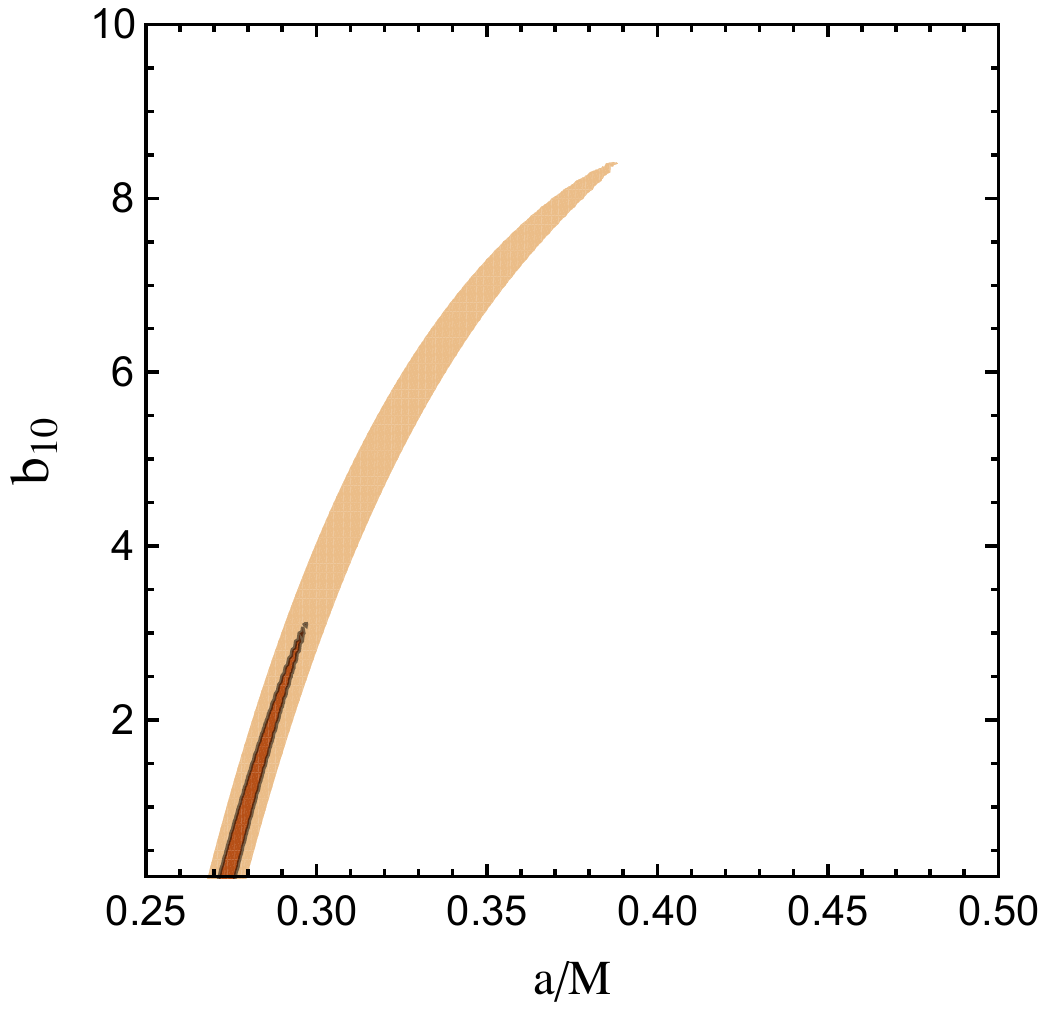}
\end{center}
\vspace{-0.3cm}
\caption{Contour levels of parameters of GB metric. Contour levels for parameters $b_2$ (top left panel), $b_4$ (top right panel), $b_5$ (bottom left panel), $b_7$ (bottom right panel). The brown, orange and lighter orange are for $1-\sigma$, $2-\sigma$ and $3-\sigma$ respectively. See the text for more details.
\label{con1}}
\end{figure*}

We first minimize $ \chi^2$ here over parameter $r$ and then fix $r$ to obtain constraint on $b_i$.\footnote{Set $r=r_{\rm ISCO}$, we may get a much worse fitting.} Figure~\ref{con1} shows the contour levels for parameters $b_2$ (top-left panel), $b_9$ (top-right panel), and $b_{10}$ (bottom panel), respectively. As seen from this Figure, $b_2$ can be well constrained to a value of $2.2^{+0.395}_{-0.523}$. The minimum of $\chi^2$ is $0.02$ and $\Delta \chi^2$ for  $1\sigma, 2\sigma$, and $3\sigma$ confidence levels are $1$, $4$, and $9$, respectively. The fitting results suggest that the BH in GRO J1655-40 may be better described by the GB metric with $b_2\neq 1$, but the significance is less than $2\sigma$ and a Kerr metric is also compatible with the data. This figure also shows degeneracies between $b_9$ and $a$ and between $b_{10}$ and $a$.

\subsection{Mock QPO observations}
\label{subsec:mock}

The QPO observations of a single object have already enabled some constraints on the GB metric. If there were QPO observations for many more objects, it may be possible to get better constraints. In this subsection, we investigate whether better constraints can be obtained from QPO measurements of multiple objects and whether the GB metric can be distinguished from the Kerr metric. To do this, we first generate $10$ Kerr BHs, each with a randomly assigned mass and spin. We also generate another mock GB BH sample, each with a randomly assigned mass and spin but a fixed $b_2=5$. For both mock samples, we calculate $\nu_\phi$, $\nu_p$, and $\nu_n$ for each sample object. We assume that these frequencies are correspondingly the QPO frequencies that can be measured with typical errors of $1\%$, similar to that of GRO J1655-40, and thus we get mock QPO observations. We adopt the Bayesian method to investigate whether whether $b_2$ can be strongly constrained as follows.

For a given set of $n$ observations $\vec{d}$, from Bayes' theorem:
\begin{equation}
\label{eq:bayes}
p(\vec{\theta}|\vec{d}) \propto p(\vec{d}|\vec{\theta})p(\vec{\theta}),
\end{equation}
where $p(\vec{\theta}|\vec{d})$ is the posterior probability distributions of the parameters $\theta$ being considered, $p(\vec{\theta})$ represent the prior information on the parameters, and $p(\vec{d}|\vec{\theta})$ is the likelihood of the observations, i.e.,
\begin{equation}
\label{eq:lnlike}
\ln p(\vec{d}|\vec{\theta}) \propto \chi^2 = \sum_{j=1}^{n} \chi^2_j.
\end{equation}
For QPO observations of each object $j$, $\chi^2_j$ can be obtained similarly as before:
\begin{equation}
\label{eq:chi2_j}
\chi^2_j = \left( \frac{\nu^j_\phi - \nu^j_{\phi_o}}{\sigma^j_{\nu_{\phi_o}}} \right)^2
         + \left( \frac{\nu^j_p - \nu^j_{p_o}}{\sigma^j_{\nu_{p_o}}} \right)^2
         + \left( \frac{\nu^j_n-\nu^j_{n_o}}{\sigma^j_{\nu_{n_o}}}\right)^2 \,,
\end{equation}
where $\nu^j_\phi$, $\nu^j_p$, and $\nu^j_n$ are the frequencies from the GB metric with given $b_i$, ($\nu^j_{\phi_o}$, $\nu^j_{p_o}$, $\nu^j_{n_o}$) and ($\sigma^j_{\nu_{\phi_o}}, \sigma^j_{\nu_{p_o}}, \sigma^j_{\nu_{n_o}}$) are the mock frequencies and its corresponding uncertainties, respectively.

\textcolor{black}{We first mock the observation data by assuming the Kerr metric,
and fit the data with the generalized GB metric. We only vary one of the
GB parameters $b_i$ each time. For each of those cases that only one $b_i$ is considered to be free, there are totally 21 parameters $\vec{\theta}=\{b_i, M_1,...M_{10}, a_1, ..., a_{10}\}$ in the model.}

In Mock data production, we also assume $r=r_{\rm ISCO}$, for simplicity.
\textcolor{black}{We sample the posterior with nested sampling algorithm using {\it dynesty} \cite{Speagle2019}. The prior for all the parameters are set to be flat with $b_i \in [0,7]$\footnote{Here we adopt a narrow range for $b_i$ as we assume that the deviation from the Kerr metric is small. In principle, a larger range of $b_i$ can be adopted, which may give a looser constraint on $b_i$ compared to the present one.}, $M_j\in[4, 11]M_\odot$ and $a_j\in[0.01,0.99]$ with $j=1, ..., 10$.}

	The result shows that $b_2$, $b_9$, and $b_{10}$ can be well constrained as expected from previous analysis shown by Figure~\ref{freq1}, \ref{freq2}, and \ref{freq3}.
Note that $b_7$ cannot be well constrained due to little frequency difference at ISCO induced by the change of $b_7$.

However, $b_4$ also cannot be well constrained though significant differences in $\nu$ can be found for cases with different $b_4$ (see Fig.~\ref{freq1}), which may be due to strong degeneracies with spin $a$ and mass $M$.

\begin{figure*}
\begin{center}
\includegraphics[width=0.49\linewidth]{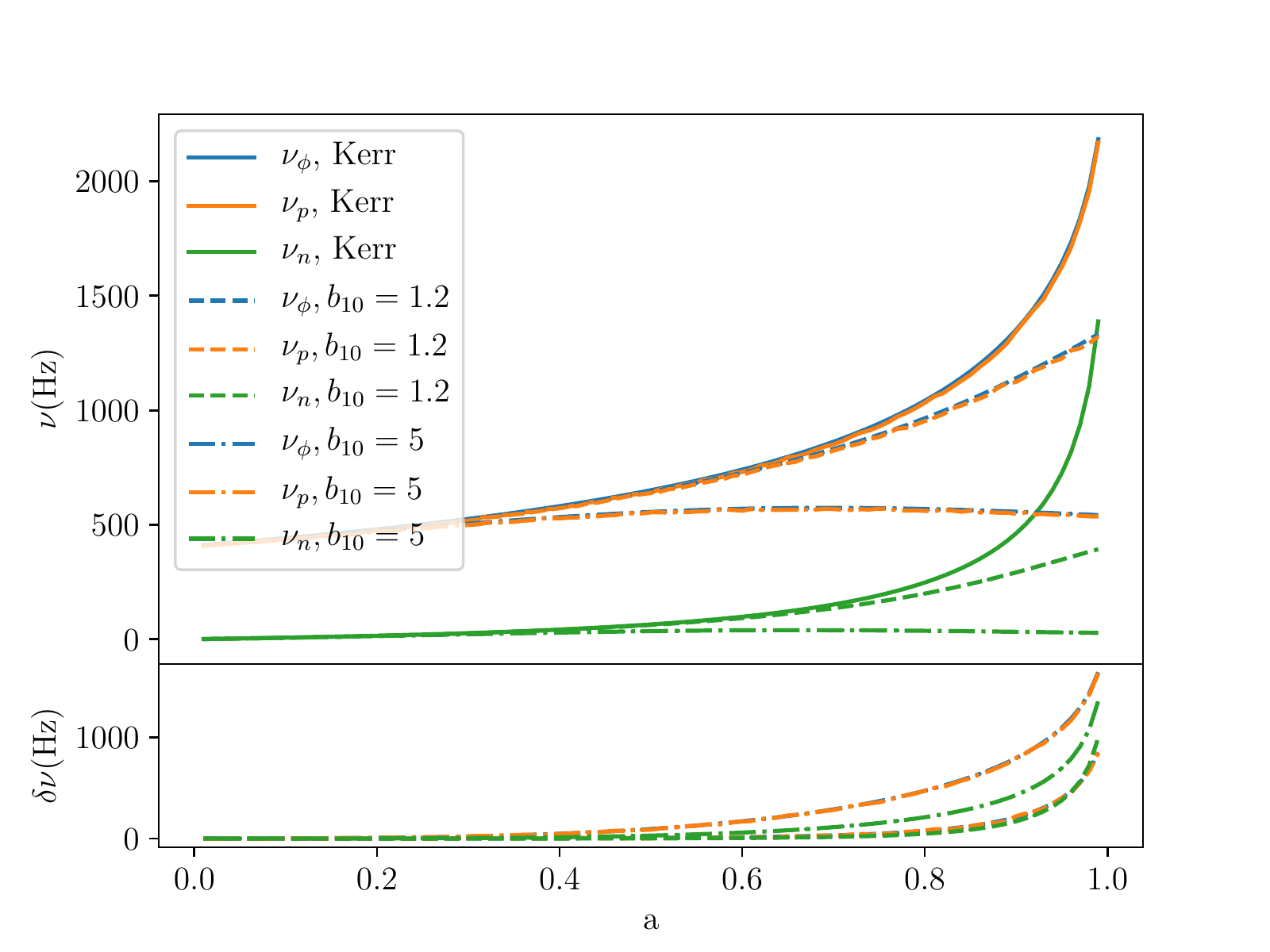}
\includegraphics[width=0.49\linewidth]{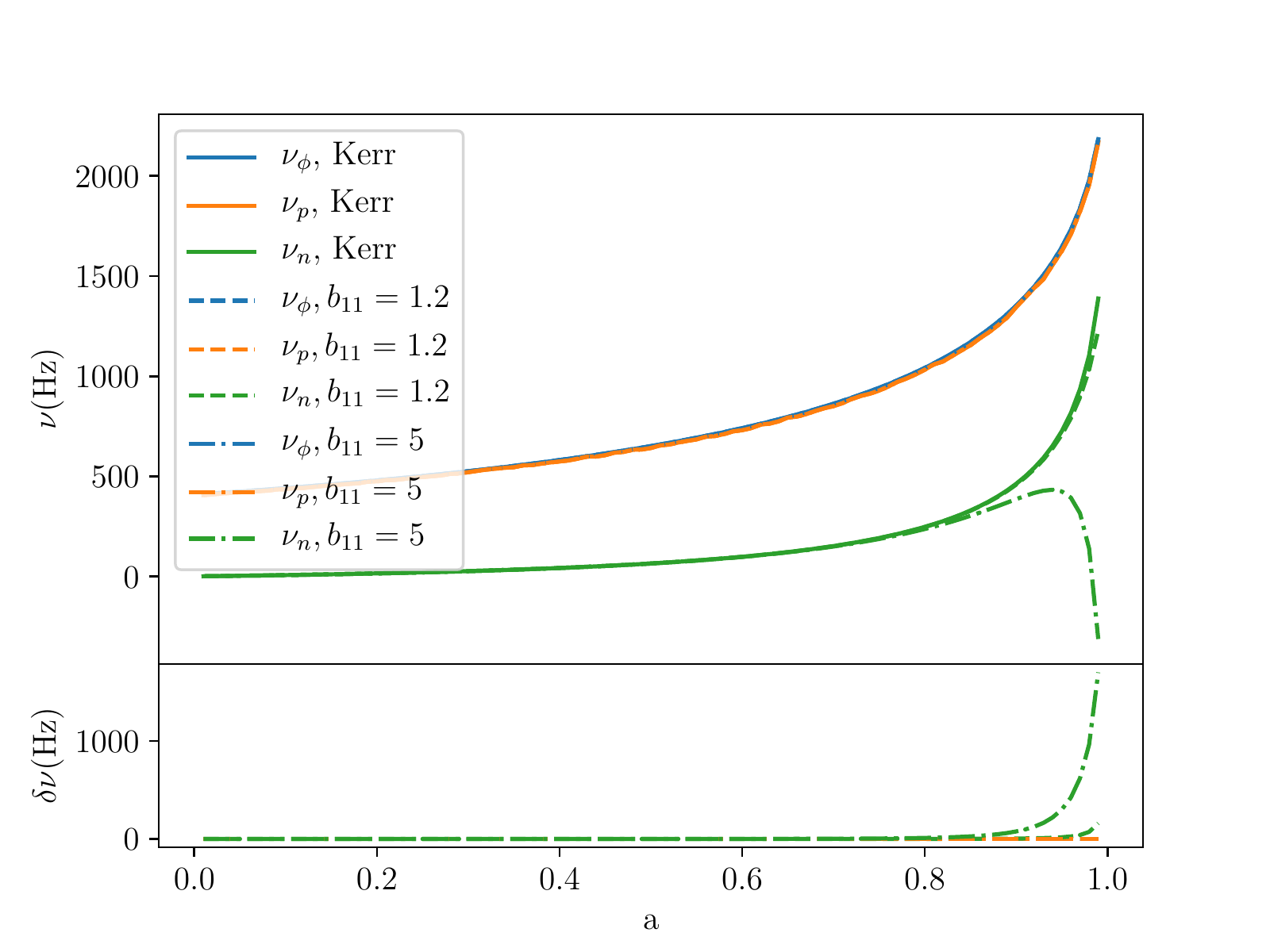}\\
\end{center}
\vspace{-0.3cm}
\caption{Dependence of the QPO frequencies at ISCO on the BH spin for a GB metric with given $b_{10}$ (left panels) and $b_{11}$ (right panels), respectively. Top panels show the resulting QPO frequencies for $b_{10}$ or $b_{11}=1, 1.2$, and $5$, respectively, while bottom panels show QPO frequency differences between the cases only with $b_{10}\neq 1$ or $b_{11}\neq 1$ and the Kerr case with all $b_i=1$.
The BH mass is set to $5.4 M_\odot$ here, the same as previous those in previous Figures. See the text for more details.
\label{b_11-a}}
\end{figure*}

The constraint on $b_{11}$ is quite good, despite the small frequencies difference induced by the change of $b_{11}$ shown in Figure \ref{freq1}.
The reason may be seen from Figure~\ref{b_11-a}, which shows the impact of $b_{10}$ (left panel) and $b_{11}$ (right panel) on QPO frequencies at ISCO as a function of the BH spin.
Apparently, the impact of $b_{10}$ gradually increases with increasing spin, while the impact of $b_{11}$ is significant only when the BH spin is high ($>0.9$). and become comparable with $b_{10}$ at high spin.
In the mock QPO data, there is only one mock BH with $a=0.93$, which is main contributor to the constraint on $b_{11}$.
As we see for other parameters like $b_5$ in Fig.~\ref{freq1} and $b_8$ in Fig.~\ref{freq2} are overlap with each other and it is hard to constrain.

	To summarize, we have shown that the GB parameters $b_2$, $b_9$, $b_{10}$, and $b_{11}$ can be well constrained if using QPO frequency measurements with a precision of $1\%$ for even only ten mock BH systems. For such a case, the reconstructed values of these $b_i$ parameters, and the mass and spin of each mock BH are listed in Table~\ref{t-obs}. If the number of measurements is larger, then better constraints may be obtained. 
One should note that the QPO frequencies are assumed to be originated at $r_{\rm ISCO}$ when generating the mock data and performing the fitting.

In realistic cases, the QPO phenomenon may not occur exactly at $r_{\rm ISCO}$, and this uncertainty may reduce the constraint power. As a comparison, in our study of GRO J1655-40 above, we minimized $\chi^2$ over $r$ to first obtain the radius for the QPO frequencies to occur. We defer detailed study on that the QPO phenomenon occurs at $r$ different from ISCO to a future work.

\begin{table*}
\caption{Inferred values of the GB parameter from the mock observations generating by assuming the Kerr metric.
}
\centering
\vspace{0.3cm}
\begin{tabular}{l|c|c|c|c|c}
\hline\hline
&Input& Best-fit\footnote{Only $b_2$ is taken as a free parameter and other $b_i$ are fixed at $1$ in the model fitting.} & Best-fit\footnote{Only $b_9$ is taken as a free parameter and other $b_i$ are fixed at $1$ in the model fitting.} & Best-fit\footnote{Only $b_{10}$ is taken as a free parameter and other $b_i$ are fixed at $1$ in the model fitting.}& Best-fit\footnote{Only $b_{11}$ is taken as a free parameter and other $b_i$ are fixed at $1$ in the model fitting.} \\
\hline
$b_2$\footnote{All other $b_i, i\neq 2$, are also set to 1.} & 1 & $1.022^{+0.042}_{-0.060}$ & $\cdots$ & $\cdots$ & $\cdots$ \\
$b_9$\footnote{All other $b_i, i\neq 9$, are also set to 1.} & 1 & $\cdots$ & $0.995^{+0.010}_{-0.676}$ & $\cdots$ & $\cdots$ \\
$b_{10}$\footnote{All other $b_i, i\neq 10$, are also set to 1.} & 1 & $\cdots$ & $\cdots$ & $0.996^{+0.004}_{-0.003} $ & $\cdots$ \\
$b_{11}$\footnote{All other $b_i, i\neq 11$, are also set to 1.} & 1 & $\cdots$ & $\cdots$ & $\cdots$ & $1.260^{+0.340}_{-0.518}$ \\
$a_{*0}$ & $0.75$ & $0.757^{+0.007}_{-0.009}$ & $0.750^{+0.006}_{-0.077}$ &  $0.753^{+0.003}_{-0.004}$ &  $0.754^{+0.003}_{-0.003}$ \\
$a_{*1}$ & $0.95$ & $0.957^{+0.018}_{-0.017}$ & $0.948^{+0.008}_{-0.152}$ &  $0.941^{+0.011}_{-0.009}$ & $0.956^{+0.012}_{-0.010}$ \\
$a_{*2}$ & $0.37$ & $0.374^{+0.003}_{-0.003}$ & $0.371^{+0.004}_{-0.010}$ &  $0.376^{+0.002}_{-0.003}$ & $0.374^{+0.002}_{-0.002}$ \\
$a_{*3}$ & $0.65$ & $0.648^{+0.004}_{-0.006}$ & $0.643^{+0.006}_{-0.050}$ &  $0.646^{+0.003}_{-0.003}$ & $0.646^{+0.003}_{-0.003}$ \\
$a_{*4}$ & $0.16$ & $0.162^{+0.001}_{-0.002}$ & $0.161^{+0.002}_{-0.002}$ &  $0.162^{+0.001}_{-0.002}$ & $0.162^{+0.001}_{-0.001}$ \\
$a_{*5}$ & $0.08$ & $0.080^{+0.001}_{-0.001}$ & $0.080^{+0.001}_{-0.001}$ &  $0.080^{+0.001}_{-0.001}$ & $0.080^{+0.001}_{-0.000}$ \\
$a_{*6}$ & $0.50$ & $0.494^{+0.004}_{-0.004}$ & $0.490^{+0.005}_{-0.022}$ &  $0.493^{+0.002}_{-0.003}$ & $0.493^{+0.003}_{-0.002}$ \\
$a_{*7}$ & $0.07$ & $0.069^{+0.001}_{-0.001}$ & $0.069^{+0.001}_{-0.001}$ &  $0.069^{+0.000}_{-0.001}$ & $0.069^{+0.000}_{-0.001}$ \\
$a_{*8}$ & $0.15$ & $0.148^{+0.052}_{-0.001}$ & $0.148^{+0.001}_{-0.001}$ &  $0.148^{+0.002}_{-0.001}$ & $0.148^{+0.002}_{-0.001}$ \\
$a_{*9}$ & $0.44$ & $0.445^{+0.003}_{-0.004}$ & $0.441^{+0.004}_{-0.017}$ &  $0.444^{+0.003}_{-0.003}$ & $0.444^{+0.003}_{-0.003}$ \\
$M_{0}/M_{\odot}$ & $5.4$ & $5.486^{+0.093}_{-0.107}$ & $5.445^{+0.051}_{-0.050}$ & $5.449^{+0.053}_{-0.051}$ & $5.452^{+0.054}_{-0.053}$ \\
$M_{1}/M_{\odot}$ & $7.9$ & $8.167^{+0.913}_{-0.626}$ & $7.830^{+0.197}_{-0.437}$ & $7.658^{+0.281}_{-0.198}$ & $8.098^{+0.593}_{-0.368}$ \\
$M_{2}/M_{\odot}$ & $9.4$ & $9.445^{+0.067}_{-0.066}$ & $9.452^{+0.066}_{-0.061}$ & $9.441^{+0.065}_{-0.064}$ & $9.441^{+0.069}_{-0.065}$ \\
$M_{3}/M_{\odot}$ & $7.8$ & $7.787^{+0.070}_{-0.111}$ & $7.758^{+0.068}_{-0.072}$ & $7.768^{+0.069}_{-0.076}$ & $7.771^{+0.068}_{-0.073}$ \\
$M_{4}/M_{\odot}$ & $6.2$ & $6.221^{+0.036}_{-0.034}$ & $6.223^{+0.035}_{-0.035}$ & $6.221^{+0.036}_{-0.035}$ & $6.221^{+0.034}_{-0.046}$ \\
$M_{5}/M_{\odot}$ & $8.4$ & $8.405^{+0.047}_{-0.045}$ & $8.411^{+0.047}_{-0.046}$ & $8.406^{+0.046}_{-0.138}$ & $8.406^{+0.046}_{-0.047}$ \\
$M_{6}/M_{\odot}$ & $8.4$ & $8.275^{+0.069}_{-0.058}$ & $8.285^{+0.070}_{-0.062}$ & $8.268^{+0.060}_{-0.055}$ & $8.268^{+0.059}_{-0.055}$ \\
$M_{7}/M_{\odot}$ & $5.3$ & $5.293^{+0.027}_{-0.027}$ & $5.295^{+0.028}_{-0.027}$ & $5.292^{+0.028}_{-0.028}$ & $5.292^{+0.028}_{-0.027}$ \\
$M_{8}/M_{\odot}$ & $7.5$ & $7.532^{+0.042}_{-0.042}$ & $7.537^{+0.044}_{-0.044}$ & $7.533^{+0.042}_{-0.042}$ & $7.532^{+0.042}_{-0.043}$ \\
$M_{9}/M_{\odot}$ & $7.2$ & $7.238^{+0.055}_{-0.051}$ & $7.248^{+0.055}_{-0.053}$ & $7.283^{+0.052}_{-0.047}$ & $7.186^{+0.051}_{-0.048}$ \\
\hline\hline
\end{tabular}
\vspace{0.1cm}
\label{t-obs}
\end{table*}

We also consider mock data by assuming the GB metric but not the Kerr ones. For each one $b_i$, we generate the mock data by setting this $b_i$ to $5$, and fit the data with the GB metric.
we only vary one $b_i$ parameter each time. 
	Similar to the Kerr case, only $b_2$, $b_9$, $b_{10}$, and $b_{11}$ can be relatively well constrained, though in most cases the constraints are not as good as those in the Kerr case. The reconstructed parameters can be found in Table~\ref{t-obs5}.
	
\begin{table*}
\centering
\caption{Values of the GB parameter inferred from the mock observations generating by assuming a GB metric with only $b_2\neq 1$ but $=5$.
}
\vspace{0.3cm}
\begin{tabular}{l|c|c|c|c|c}
\hline\hline
& Input & Best-fit\footnote{Only $b_2$ is taken as a free parameter and other $b_i$ are fixed at $1$ in the model fitting.} & Best-fit\footnote{Only $b_9$ is taken as a free parameter and other $b_i$ are fixed at $1$ in the model fitting.} & Best-fit\footnote{Only $b_{10}$ is taken as a free parameter and other $b_i$ are fixed at $1$ in the model fitting.} & Best-fit\footnote{Only $b_{11}$ is taken as a free parameter and other $b_i$ are fixed at $1$ in the model fitting.}\\
\hline
$b_2$\footnote{All other $b_i, i\neq 2$, are set to $1$.} & 5 & $4.993^{+0.006}_{-0.010}$ & $\cdots$ & $\cdots$ & $\cdots$ \\
$b_9$\footnote{All other $b_i, i\neq 9$, are set to $1$.} & 5 & $\cdots$ & $4.934^{+0.101}_{-0.065}$ & $\cdots$ & $\cdots$ \\
$b_{10}$\footnote{All other $b_i, i\neq 10$, are set to $1$.} & 5 & $\cdots$ & $\cdots$ & $5.091^{+0.063}_{-0.245}$ & $\cdots$ \\
$b_{11}$\footnote{All other $b_i, i\neq 11$, are set to $1$.} & 5 & $\cdots$ & $\cdots$ & $\cdots$ & $5.709^{+0.946}_{-2.371}$  \\
$a_{*0}$ & $0.75$ & $0.751^{+0.005}_{-0.003}$ & $0.550^{+0.250}_{-0.057}$ & $0.592^{+0.162}_{-0.078}$ & $0.749^{+0.007}_{-0.006}$   \\
$a_{*1}$ & $0.95$ & $0.949^{+0.002}_{-0.001}$ & $0.964^{+0.019}_{-0.020}$ & $0.408^{+0.517}_{-0.013}$ & $0.865^{+0.032}_{-0.024}$   \\
$a_{*2}$ & $0.37$ & $0.396^{+0.037}_{-0.032}$ & $0.910^{+0.033}_{-0.544}$ & $0.952^{+0.029}_{-0.578}$ & $0.370^{+0.002}_{-0.002}$   \\
$a_{*3}$ & $0.65$ & $0.650^{+0.007}_{-0.003}$ & $0.635^{+0.066}_{-0.064}$ & $0.630^{+0.085}_{-0.083}$ & $0.650^{+0.003}_{-0.003}$   \\
$a_{*4}$ & $0.16$ & $0.160^{+0.002}_{-0.002}$ & $0.164^{+0.002}_{-0.002}$ & $0.161^{+0.001}_{-0.001}$ & $0.158^{+0.001}_{-0.001}$   \\
$a_{*5}$ & $0.08$ & $0.080^{+0.001}_{-0.001}$ & $0.080^{+0.001}_{-0.001}$ & $0.079^{+0.001}_{-0.001}$ & $0.078^{+0.001}_{-0.001}$   \\
$a_{*6}$ & $0.50$ & $0.248^{+0.213}_{-0.005}$ & $0.520^{+0.273}_{-0.027}$ & $0.515^{+0.314}_{-0.042}$ & $0.513^{+0.003}_{-0.004}$   \\
$a_{*7}$ & $0.07$ & $0.070^{+0.530}_{-0.000}$ & $0.068^{+0.001}_{-0.001}$ & $0.068^{+0.001}_{-0.001}$ & $0.069^{+0.001}_{-0.001}$   \\
$a_{*8}$ & $0.15$ & $0.151^{+0.001}_{-0.002}$ & $0.150^{+0.001}_{-0.001}$ & $0.149^{+0.001}_{-0.001}$ & $0.153^{+0.001}_{-0.001}$   \\
$a_{*9}$ & $0.44$ & $0.400^{+0.044}_{-0.006}$ & $0.439^{+0.403}_{-0.012}$ & $0.859^{+0.054}_{-0.404}$ & $0.445^{+0.003}_{-0.003}$   \\
$M_{0}/M_{\odot}$ & $5.4$ & $5.442^{+0.060}_{-0.043}$ & $5.431^{+0.052}_{-0.048}$ & $5.238^{+0.049}_{-0.047}$ & $5.336^{+0.104}_{-0.070}$  \\
$M_{1}/M_{\odot}$ & $7.9$ & $7.904^{+0.078}_{-0.068}$ & $7.873^{+0.070}_{-0.054}$ & $7.677^{+0.174}_{-0.069}$ & $5.820^{+0.604}_{-0.349}$  \\
$M_{2}/M_{\odot}$ & $9.4$ & $9.670^{+0.412}_{-0.308}$ & $9.289^{+0.165}_{-0.101}$ & $9.797^{+0.091}_{-0.239}$ & $9.441^{+0.057}_{-0.055}$  \\
$M_{3}/M_{\odot}$ & $7.8$ & $7.770^{+0.070}_{-0.062}$ & $7.887^{+0.075}_{-0.068}$ & $7.746^{+0.065}_{-0.056}$ & $7.802^{+0.058}_{-0.065}$  \\
$M_{4}/M_{\odot}$ & $6.2$ & $6.230^{+0.044}_{-0.039}$ & $6.314^{+0.038}_{-0.038}$ & $6.175^{+0.036}_{-0.036}$ & $6.138^{+0.034}_{-0.033}$  \\
$M_{5}/M_{\odot}$ & $8.4$ & $8.296^{+0.045}_{-0.045}$ & $8.340^{+0.048}_{-0.048}$ & $8.256^{+0.045}_{-0.043}$ & $8.357^{+0.045}_{-0.043}$  \\
$M_{6}/M_{\odot}$ & $8.4$ & $6.743^{+1.665}_{-0.066}$ & $8.333^{+0.076}_{-0.079}$ & $8.354^{+0.099}_{-0.097}$ & $8.623^{+0.064}_{-0.074}$  \\
$M_{7}/M_{\odot}$ & $5.3$ & $5.332^{+3.766}_{-0.033}$ & $5.323^{+0.030}_{-0.029}$ & $5.229^{+0.028}_{-0.028}$ & $5.274^{+0.027}_{-0.027}$  \\
$M_{8}/M_{\odot}$ & $7.5$ & $7.546^{+0.044}_{-0.044}$ & $7.468^{+0.044}_{-0.042}$ & $7.381^{+0.042}_{-0.041}$ & $7.620^{+0.043}_{-0.043}$  \\
$M_{9}/M_{\odot}$ & $7.2$ & $6.937^{+0.332}_{-0.306}$ & $7.218^{+0.062}_{-0.119}$ & $7.499^{+0.109}_{-0.122}$ & $7.303^{+0.061}_{-0.054}$  \\
\hline\hline
\end{tabular}
\vspace{0.1cm}
\label{t-obs5}
\end{table*}

We may compare these results with those using BH shadow and iron line methods as follows. BH shadow \cite{p15} shows very weak dependent on the parameters $b_2$, $b_8$, $b_9$ and $b_{10}$. Parameters $b_4$, $b_5$, $b_7$ and $b_{11}$ do not produce any impact on shadow boundary and thus cannot be constrained.
According to the iron line studies in \cite{ghasemi}, all GB parameters except $b_{11}$ can be constrained with near future X-ray mission, though $b_5$ and $b_8$ are relatively more difficult to constrain.
Different methods may be complementary to each other and put constraints on different deformation GB parameters, therefore, they can be combined together to help break the degeneracies and finally pin down the metric of black holes.


\section{Discussions}\label{Discussion}

In this paper, we simply assume that QPOs are related to the fundamental frequencies at the same radius in the geodesic models for QPOs \cite{rpm}. However, we note that the QPOs may not occur at the same radius even they are indeed determined by those fundamental frequencies as that in the geodesic models \cite{StuchlikApJ}. It is also possible that the QPOs may be related to other frequencies, such as that proposed in the epicycle resonance (ER) model \cite{Torok11}. Therefore, our demonstrations on the constraining power of QPOs on the metric parameters $b_i$ are valid when the geodesic models correctly explain the QPO phenomena and our results may be model dependent. The study presented in our paper may be extended to those cases that the QPOs are due to other frequencies, such as those in the ER model, which is beyond the scope of this paper as currently it is still clear which model can explain the QPOs the best. With many QPO measurements in future, one could combine different QPO models and BH metrics to obtain strong constraints on both the QPO model and the metric parameters.

We also note that there are various works in literature on constraining alternative metrics for black holes. Below we give a short summary for them and compare some of them with ours presented in this paper. 

Reference~\cite{q1} checked braneworld Kerr BH when bulk-space influence is described by a single, brany tidal charge parameter, $b$. Similar to Kerr-Newman solution in GR in which the square of the electric charge $Q^2$ is replaced by a tidal charge $b$. The behavior of radial and vertical epicyclic frequencies is qualitatively similar to Kerr and brany Kerr BHs. But there are strong differences in the case of naked singularities. The vertical epicyclic frequency could be even lower than radial one. In the structure of radial profile the number of local extrema could be higher in comparison with standard Kerr naked singularity. Also, the radial epicyclic frequency has no zero point for some special family of brany naked singularity.
Reference~\cite{q2} introduced the Jahannsen and Psaltis (JP) metric and studied their QPOs. In their quasi Kerr space time, quadrupole moment is a free parameter in addition to Mass and spin. They show that for moderate spin, the Keplerian frequency is independent of small deviations of the quadrupole moment from Kerr value. They also showed that the epicyclic frequencies shows significant variations.
Reference~\cite{q4} study QPOs in the space time of rotating braneworld gravity. Such a BH carries a tidal charge as imprint of the extra fifth dimension. The metric contains $\beta$ parameter as tidal charge. They found that large enough value of the positive tidal charge is not supported by observations of high frequency QPOs. But for large enough negative tidal charge, the braneworld BH are similar to the Kerr case. They also showed that over-rotating braneworld and extreme Kerr BH can not be distinguished in the high frequency QPOs observations.
Reference~\cite{q5} considered Bardeen BH metric where parameter $g$ in the metric can be introduced as the magnetic charge of a non-linear electromagnetic field. He also considered JP metric where metric has an infinite number of deformation parameter $\epsilon_k$. He fixes the mass and found the spin parameter $a/M = 0.279$ and $g/M < 0.56$ at the $68\%$ C.L.. For second metric he found $a/M = 0.27$ and $\epsilon_3 = 0.5$ at the $68\%$ C.L..
In the reference~\cite{q6}, they studied Einstein-Dilaton-Gauss-Bonnet (EDGB) theory with QPOs. $\alpha /M^2$ characterizes the theory. They considered low spinning BH in this work and showed the LOFT observation can put constrain on the parameter $\alpha /M^2$ of EDGB.
Reference~\cite{q7} considered Konoplya and Zhikendao metric. The metric is obtained by deforming the Kerr metric by adding a static deformation to mass. They found constraint on spin parameter and deformation parameter $\delta r / r_{Kerr}$.\\
If we compare our work with these above mentioned work, we measure QPO frequencies and use MCMC method in addition to directly check data with observations.

\section{Conclusions \label{summary}}

In this paper, we investigate whether QPOs can be used to constrain the parameters of the  GB metric introduced in Ref.~\cite{p15} and possible deviations from the Kerr one.

We adopt the general relativistic precession model to relate the QPO frequencies to those fundamental ones, including the periastron and nodal precession frequencies, of test particle at the inner edge of accretion disks around black holes.
We find that at least some of the QPO frequencies resulting from the GB metric with $b_2$, $b_4$, $b_9$, $b_{10}$, or $b_{11}$ significantly deviating from $1$ can be substantially different from those from the Kerr case, though there are no such differences for other $b_i$s.

Considering the case of GRO J1655-40 with three accurately measured QPO frequencies, we find that 
the GB parameter $b_2$ is required to be $2.2^{+0.395}_{-0.523}$ but other GB parameters cannot be well constrained. Although such a constraint on $b_2$ seems to indicate that a GB metric fits to the QPO data better, but the significance is smaller than $2$-$\sigma$.
By generating mock samples of $10$ black hole systems and each black hole with three QPO frequency measurements, we use the nested algorithm to reconstruct the input $b$ parameter(s) by using these mock samples. We find that $b_2$, $b_9$, $b_{10}$, and $b_{11}$ can be well reconstructed, while $b_5$, $b_7$, and $b_8$ cannot, as expected. However, $b_4$ cannot also be well constrained due to parameter degeneracies among $b_4$, $M$, and $a$. The results presented in this paper suggest that strong constraints on a few GB parameters can be obtained by using the QPO frequency measurements of only about ten or more black hole systems.


{\bf Acknowledgments}
This work is partly supported by the National Key Program for Science and Technology Research and Development (Grant No.~2016YFA0400704), the National Natural Science Foundation of China under grant
No.~11690024 and~11390372, and the Strategic Priority Program of the Chinese Academy of Sciences (Grant
No.~XDB 23040100). M. G.-N. also acknowledge support from the China Postdoctoral Science Foundation, Grant No.~2017LH021.



\begin{thebibliography}{}


\bibitem{einstein1916}
Einstein, A.\ 1916, Annalen der Physik, 354, 769

\bibitem{t1}
  C.~M.~Will,
   ``The Confrontation between General Relativity and Experiment,''
 \href{http://dx.doi.org/10.12942/lrr-2014-4}{Living Rev.\ Rel.\  {\bf 17} (2014) 4}
  [arXiv:1403.7377 [gr-qc]].

   \bibitem{t2}
{Will} C~M 1993 {\em {Theory and Experiment in Gravitational Physics}\/}
  (Cambridge University Press) ISBN 0521439736


\bibitem{t3}
Stairs I~H 2003,
``Testing General Relativity with Pulsar Timing,"
\href{http://dx.doi.org/10.12942/lrr-2003-5}{ {Living Reviews in Relativity\/} {\bf 6} }

\bibitem{t4}
Wex N 2014 {``Testing Relativistic Gravity with Radio Pulsars"} {Frontiers in
  Relativistic Celestial Mechanics\/,} vol~1 ed Kopeikin S (De Gruyter) ISBN
  9783110345667 [arXiv:1402.5594]


 \bibitem{k1}
  R.~P.~Kerr,
  ``Gravitational field of a spinning mass as an example of algebraically special metrics,''
  \href{http://dx.doi.org/10.1103/PhysRevLett.11.237}{Phys.\ Rev.\ Lett.\  {\bf 11} (1963) 237.}


\bibitem{k2}
  R.~P.~Kerr,
  ``Gravitational collapse and rotation,''


   \bibitem{g1}
  J.~R.~Gair, M.~Vallisneri, S.~L.~Larson and J.~G.~Baker,
  ``Testing General Relativity with Low-Frequency, Space-Based Gravitational-Wave Detectors,''
  \href{http://dx.doi.org/10.12942/lrr-2013-7}{Living Rev.\ Rel.\  {\bf 16}, 7 (2013)}
  [arXiv:1212.5575 [gr-qc]].


  \bibitem{g2}
  N.~Yunes and X.~Siemens,
 ``Gravitational-Wave Tests of General Relativity with Ground-Based Detectors and Pulsar Timing-Arrays,''
  \href{http://dx.doi.org/10.12942/lrr-2013-9}{Living Rev.\ Rel.\  {\bf 16}, 9 (2013)}
  [arXiv:1304.3473 [gr-qc]].


  \bibitem{e1}
  D.~Psaltis,
  ``Probes and Tests of Strong-Field Gravity with Observations in the Electromagnetic Spectrum,''
  \href{http://dx.doi.org/10.12942/lrr-2008-9}{Living Rev.\ Rel.\  {\bf 11}, 9 (2008)}
  [arXiv:0806.1531 [astro-ph]].


\bibitem{e2}
  C.~Bambi,
  ``Testing black hole candidates with electromagnetic radiation,''
  Rev.\ Mod.\ Phys.\  {\bf 89}, no. 2, 025001 (2017)
  doi:10.1103/RevModPhys.89.025001
  [arXiv:1509.03884 [gr-qc]].


\bibitem{p1}
Manko, V.~S., \& Novikov, I.~D.\ 1992, Classical and Quantum Gravity, 9, 2477


 \bibitem{p2}
  K.~Glampedakis and S.~Babak,
  ``Mapping spacetimes with LISA: Inspiral of a test-body in a `quasi-Kerr' field,''
\href{http://dx.doi.org/10.1088/0264-9381/23/12/013}{Class.\ Quant.\ Grav.\  {\bf 23}, 4167 (2006)}
  [gr-qc/0510057].


\bibitem{p3}
  S.~J.~Vigeland and S.~A.~Hughes,
 ``Spacetime and orbits of bumpy black holes,''
  \href{http://dx.doi.org/10.1103/PhysRevD.81.024030}{Phys.\ Rev.\ D {\bf 81}, 024030 (2010)}
  [arXiv:0911.1756 [gr-qc]].


\bibitem{p4}
  S.~J.~Vigeland,
  ``Multipole moments of bumpy black holes,''
  \href{http://dx.doi.org/10.1103/PhysRevD.82.104041}{Phys.\ Rev.\ D {\bf 82}, 104041 (2010)}
  doi:10.1103/PhysRevD.82.104041
  [arXiv:1008.1278 [gr-qc]].


\bibitem{p5}
  S.~Vigeland, N.~Yunes and L.~Stein,
  ``Bumpy Black Holes in Alternate Theories of Gravity,''
  \href{http://dx.doi.org/10.1103/PhysRevD.83.104027}{Phys.\ Rev.\ D {\bf 83}, 104027 (2011)}


  \bibitem{p6}
  T.~Johannsen and D.~Psaltis,
  ``A Metric for Rapidly Spinning Black Holes Suitable for Strong-Field Tests of the No-Hair Theorem,''
  \href{http://dx.doi.org/10.1103/PhysRevD.83.124015}{Phys.\ Rev.\ D {\bf 83}, 124015 (2011)}
  [arXiv:1105.3191 [gr-qc]].

\bibitem{p7}
  V.~Cardoso, P.~Pani and J.~Rico,
 ``On generic parametrizations of spinning black-hole geometries,''
 \href{http://dx.doi.org/10.1103/PhysRevD.89.064007}{ Phys.\ Rev.\ D {\bf 89}, 064007 (2014)}
  doi:10.1103/PhysRevD.89.064007
  [arXiv:1401.0528 [gr-qc]].


    \bibitem{p8}
  L.~Rezzolla and A.~Zhidenko,
 ``New parametrization for spherically symmetric black holes in metric theories of gravity,''
  \href{http://dx.doi.org/10.1103/PhysRevD.90.084009}{Phys.\ Rev.\ D {\bf 90}, no. 8, 084009 (2014)}
  [arXiv:1407.3086 [gr-qc]].


  \bibitem{p9}
  N.~Lin, N.~Tsukamoto, M.~Ghasemi-Nodehi and C.~Bambi,
 ``A parametrization to test black hole candidates with the spectrum of thin disks,''
  \href{http://dx.doi.org/10.1140/epjc/s10052-015-3837-3}{Eur.\ Phys.\ J.\ C {\bf 75}, no. 12, 599 (2015)},
  [arXiv:1512.00724 [gr-qc]].


\bibitem{p10}
  R.~Konoplya, L.~Rezzolla and A.~Zhidenko,
 ``General parametrization of axisymmetric black holes in metric theories of gravity,''
  \href{http://dx.doi.org/10.1103/PhysRevD.93.064015}{Phys.\ Rev.\ D {\bf 93}, no. 6, 064015 (2016)}
  [arXiv:1602.02378 [gr-qc]].


 \bibitem{p11}
  C.~Bambi, J.~Jiang and J.~F.~Steiner,
 ``Testing the no-hair theorem with the continuum-fitting and the iron line methods: a short review,''
  \href{http://dx.doi.org/10.1088/0264-9381/33/6/064001}{Class.\ Quant.\ Grav.\  {\bf 33}, no. 6, 064001 (2016)}
  [arXiv:1511.07587 [gr-qc]].


   \bibitem{p12}
  K.~Yagi and L.~C.~Stein,
 ``Black Hole Based Tests of General Relativity,''
  \href{http://dx.doi.org/10.1088/0264-9381/33/5/054001}{Class.\ Quant.\ Grav.\  {\bf 33}, no. 5, 054001 (2016)}
    [arXiv:1602.02413 [gr-qc]].


  \bibitem{p13}
  T.~Johannsen,
 ``Testing the No-Hair Theorem with Observations of Black Holes in the Electromagnetic Spectrum,''
  \href{http://dx.doi.org/10.1088/0264-9381/33/12/124001}{Class.\ Quant.\ Grav.\  {\bf 33}, no. 12, 124001 (2016)}
  [arXiv:1602.07694 [astro-ph.HE]].


\bibitem{p14}
  K.~Yagi and L.~C.~Stein,
 ``Black Hole Based Tests of General Relativity,''
  Class.\ Quant.\ Grav.\  {\bf 33}, 054001 (2016)
  [arXiv:1602.02413 [gr-qc]].


 \bibitem{p15}
  M.~Ghasemi-Nodehi and C.~Bambi,
 ``Note on a new parametrization for testing the Kerr metric,''
 \href{http://dx.doi.org/10.1140/epjc/s10052-016-4137-2}{Eur.\ Phys.\ J.\ C {\bf 76}, no. 5, 290 (2016)},
  [arXiv:1604.07032 [gr-qc]].


   \bibitem{p16}
  V.~Cardoso and L.~Gualtieri,
  ``Testing the black hole `no-hair' hypothesis,''
  \href{http://dx.doi.org/10.1088/0264-9381/33/17/174001}{Class.\ Quant.\ Grav.\  {\bf 33}, no. 17, 174001 (2016)}
  [arXiv:1607.03133 [gr-qc]].


 \bibitem{r1}
  C.~Bambi,
 ``Testing the Kerr black hole hypothesis,''
  Mod.\ Phys.\ Lett.\ A {\bf 26}, 2453 (2011)
  [arXiv:1109.4256 [gr-qc]];


 \bibitem{r2}
  C.~Bambi,
  ``Testing the space-time geometry around black hole candidates with the available radio and X-ray data,''
  Astron.\ Rev.\  {\bf 8}, 4 (2013)
  [arXiv:1301.0361 [gr-qc]].

  \bibitem{ka}
  A.~C.~Fabian, M.~J.~Rees, L.~Stella and N.~E.~White,
  ``X-ray fluorescence from the inner disc in Cygnus X-1,''
  Mon.\ Not.\ Roy.\ Astron.\ Soc.\  {\bf 238}, 729 (1989);
  A.~C.~Fabian, K.~Iwasawa, C.~S.~Reynolds and A.~J.~Young,
  ``Broad iron lines in active galactic nuclei,''
  Publ.\ Astron.\ Soc.\ Pac.\  {\bf 112}, 1145 (2000)
  [astro-ph/0004366];
  C.~S.~Reynolds and M.~A.~Nowak,
  ``Fluorescent iron lines as a probe of astrophysical black hole systems,''
  Phys.\ Rept.\  {\bf 377}, 389 (2003)
  [astro-ph/0212065].


\bibitem{cfm}
  S.~N.~Zhang, W.~Cui and W.~Chen,
  ``Black hole spin in X-ray binaries: Observational consequences,''
  Astrophys.\ J.\  {\bf 482}, L155 (1997)
  [astro-ph/9704072];
  L.~-X.~Li, E.~R.~Zimmerman, R.~Narayan and J.~E.~McClintock,
  ``Multi-temperature blackbody spectrum of a thin accretion disk around a Kerr black hole: Model computations and comparison with observations,''
  Astrophys.\ J.\ Suppl.\  {\bf 157}, 335 (2005)
  [astro-ph/0411583];
  J.~E.~McClintock {\it et al.},
  ``Measuring the Spins of Accreting Black Holes,''
  Class.\ Quant.\ Grav.\  {\bf 28}, 114009 (2011)
  [arXiv:1101.0811 [astro-ph.HE]].

\bibitem{evh}
 http://www.eventhorizontelescope.org/science/index.html

 \bibitem{evh1}
  T.~Johannsen, C.~Wang, A.~E.~Broderick, S.~S.~Doeleman, V.~L.~Fish, A.~Loeb and D.~Psaltis,
  ``Testing General Relativity with Accretion-Flow Imaging of Sgr A*,''
  Phys.\ Rev.\ Lett.\  {\bf 117}, no. 9, 091101 (2016)
  [arXiv:1608.03593 [astro-ph.HE]].

\bibitem{rpm}
  L.~Stella and M.~Vietri,
  ``Lense-Thirring precession and QPOS in low mass x-ray binaries,''
  Astrophys.\ J.\  {\bf 492}, L59 (1998)
  [astro-ph/9709085];
  L.~Stella and M.~Vietri,
  ``Khz quasi periodic oscillations in low mass x-ray binaries as probes of general relativity in the strong field regime,''
  Phys.\ Rev.\ Lett.\  {\bf 82}, 17 (1999)
  [astro-ph/9812124];
  L.~Stella, M.~Vietri and S.~Morsink,
  ``Correlations in the qpo frequencies of low mass x-ray binaries and the relativistic precession model,''
  Astrophys.\ J.\  {\bf 524}, L63 (1999)
  [astro-ph/9907346].

\bibitem{dm}
  C.~A.~Perez, A.~S.~Silbergleit, R.~V.~Wagoner and D.~E.~Lehr,
  ``Relativistic diskoseismology. 1. Analytical results for 'gravity modes',''
  Astrophys.\ J.\  {\bf 476}, 589 (1997)
  [astro-ph/9601146];
  A.~S.~Silbergleit, R.~V.~Wagoner and M.~Ortega-Rodriguez,
  ``Relativistic diskoseismology. 2. Analytical results for C modes,''
  Astrophys.\ J.\  {\bf 548}, 335 (2001)
  [astro-ph/0004114];
  S.~Kato,
  ``Basic Properties of Thin-Disk Oscillations,''
  Publ.\ Astron.\ Soc.\ Jap.\  {\bf 53} 1 (2001).

\bibitem{rm}
  M.~A.~Abramowicz and W.~Kluzniak,
  ``A Precise determination of angular momentum in the black hole candidate GRO J1655-40,''
  Astron.\ Astrophys.\  {\bf 374}, L19 (2001)
  [astro-ph/0105077];
  M.~A.~Abramowicz, V.~Karas, W.~Kluzniak, W.~H.~Lee and P.~Rebusco,
  ``Non-Linear Resonance in Nearly Geodesic Motion in Low-Mass X-Ray Binaries,''
  Publ.\ Astron.\ Soc.\ Jap.\  {\bf 55} 467 (2003);
  G.~Torok, M.~A.~Abramowicz, W.~Kluzniak and Z.~Stuchlik,
  ``The orbital resonance model for twin peak kHz quasi periodic oscillations in microquasars,''
  Astron.\ Astrophys.\  {\bf 436}, 1 (2005).

\bibitem{rezz}
  L.~Rezzolla, S.~'i.~Yoshida, T.~J.~Maccarone and O.~Zanotti,
  ``A New simple model for high frequency quasi periodic oscillations in black hole candidates,''
  Mon.\ Not.\ Roy.\ Astron.\ Soc.\  {\bf 344}, L37 (2003)
  [astro-ph/0307487];
  J.~D.~Schnittman and L.~Rezzolla,
  ``Quasi-periodic oscillations in the x-ray light curves from relativistic tori,''
  Astrophys.\ J.\  {\bf 637}, L113 (2006)
  [astro-ph/0506702].


\bibitem{q1}
  Z.~Stuchlik and A.~Kotrlova,
  ``Orbital resonances in discs around braneworld Kerr black holes,''
  Gen.\ Rel.\ Grav.\  {\bf 41}, 1305 (2009)
  [arXiv:0812.5066 [astro-ph]].


  \bibitem{q2}
  T.~Johannsen and D.~Psaltis,
  ``Testing the No-Hair Theorem with Observations in the Electromagnetic Spectrum. III. Quasi-Periodic Variability,''
  Astrophys.\ J.\  {\bf 726}, 11 (2011)
  [arXiv:1010.1000 [astro-ph.HE]].


  \bibitem{q3}
  C.~Bambi,
  ``Probing the space-time geometry around black hole candidates with the resonance models for high-frequency QPOs and comparison with the continuum-fitting method,''
  JCAP {\bf 1209}, 014 (2012)
  [arXiv:1205.6348 [gr-qc]].


\bibitem{q4}
  A.~N.~Aliev, G.~D.~Esmer and P.~Talazan,
  ``Strong Gravity Effects of Rotating Black Holes: Quasiperiodic Oscillations,''
  Class.\ Quant.\ Grav.\  {\bf 30}, 045010 (2013)
  [arXiv:1205.2838 [gr-qc]].

  \bibitem{q5}
  C.~Bambi,
  ``Testing the nature of the black hole candidate in GRO J1655-40 with the relativistic precession model,''
  Eur.\ Phys.\ J.\ C {\bf 75}, no. 4, 162 (2015)
  [arXiv:1312.2228 [gr-qc]].

  \bibitem{q6}
  A.~Maselli, L.~Gualtieri, P.~Pani, L.~Stella and V.~Ferrari,
  ``Testing Gravity with Quasi Periodic Oscillations from accreting Black Holes: the Case of the Einstein-Dilaton-Gauss-Bonnet Theory,''
  Astrophys.\ J.\  {\bf 801}, no. 2, 115 (2015)
  [arXiv:1412.3473 [astro-ph.HE]].

  \bibitem{q7}
  C.~Bambi and S.~Nampalliwar,
  ``Quasi-periodic oscillations as a tool for testing the Kerr metric: A comparison with gravitational waves and iron line,''
  EPL {\bf 116}, no. 3, 30006 (2016)
  [arXiv:1604.02643 [gr-qc]].



  \bibitem{ghasemi}
  M.~Ghasemi-Nodehi and C.~Bambi,
  ``Constraining the Kerr parameters via X-ray reflection spectroscopy,''
  Phys.\ Rev.\ D {\bf 94}, no. 10, 104062 (2016)
  [arXiv:1610.08791 [gr-qc]].

  \bibitem{bookbambi}
  C.~Bambi,
  ``Black Holes: A Laboratory for Testing Strong Gravity,''


\bibitem{Motta1}
  S.~E.~Motta, T.~M.~Belloni, L.~Stella, T.~Muñoz-Darias and R.~Fender,
  ``Precise mass and spin measurements for a stellar-mass black hole through X-ray timing: the case of GRO J1655-40,''
  Mon.\ Not.\ Roy.\ Astron.\ Soc.\  {\bf 437}, no. 3, 2554 (2014)
  [arXiv:1309.3652 [astro-ph.HE]].


\bibitem{Motta2}
  S.~E.~Motta, T.~Muñoz-Darias, A.~Sanna, R.~Fender, T.~Belloni and L.~Stella,
  ``Black hole spin measurements through the relativistic precession model: XTE J1550-564,''
  Mon.\ Not.\ Roy.\ Astron.\ Soc.\  {\bf 439}, 65 (2014)
  [arXiv:1312.3114 [astro-ph.HE]].

 \bibitem{gro2}
  P.~Casella, T.~Belloni and L.~Stella,
  ``The ABC of low-frequency quasi-periodic oscillations in black-hole candidates: Analogies with Z-sources,''
  Astrophys.\ J.\  {\bf 629}, 403 (2005)
  doi:10.1086/431174
  [astro-ph/0504318].

  \bibitem{opt}
  M.~E.~Beer and P.~Podsiadlowski,
  ``The quiescent light curve and evolutionary state of gro j1655-40,''
  Mon.\ Not.\ Roy.\ Astron.\ Soc.\  {\bf 331}, 351 (2002)
  [astro-ph/0109136].

\bibitem{Speagle2019} Speagle, J.~S.\ 2019, arXiv e-prints, arXiv:1904.02180

\bibitem{StuchlikApJ} Z.~Stuchl{\'\i}k,  \& M.~Kolo{\v{s}}, ``Controversy of the GRO J1655-40 Black Hole Mass and Spin Estimates and Its Possible Solutions'',
Astrophys.\ J.\ {\bf 825}, 13 (2016)

\bibitem{Torok11}  G.~T{\"o}r{\"o}k, Kotrlov{\'a}, A., E.~{\v{S}}r{\'a}mkov{\'a},  et al.\
``Confronting the models of 3:2 quasiperiodic oscillations with the rapid spin of the microquasar GRS 1915+105'', Astron. \& Astrophy. {\bf 531}, A59 (2011)

\end{thebibliography}
\end{document}